\PassOptionsToPackage{unicode}{hyperref}
\PassOptionsToPackage{hyphens}{url}
\PassOptionsToPackage{dvipsnames,svgnames,x11names}{xcolor}
\documentclass[
  12pt]{article}

\usepackage{amsmath,amssymb} 
\newtheorem{theorem}{Theorem} 
\newtheorem{definition}{Definition}

\usepackage{algorithm}
\usepackage{algpseudocode}
\usepackage{comment}

\usepackage{iftex}
\ifPDFTeX
  \usepackage[T1]{fontenc}
  \usepackage[utf8]{inputenc}
  \usepackage{textcomp} 
\else 
  \usepackage{unicode-math}
  \defaultfontfeatures{Scale=MatchLowercase}
  \defaultfontfeatures[\rmfamily]{Ligatures=TeX,Scale=1}
\fi
\usepackage{lmodern}
\ifPDFTeX\else  
\fi
\IfFileExists{upquote.sty}{\usepackage{upquote}}{}
\IfFileExists{microtype.sty}{
  \usepackage[]{microtype}
  \UseMicrotypeSet[protrusion]{basicmath} 
}{}
\makeatletter
\@ifundefined{KOMAClassName}{
  \IfFileExists{parskip.sty}{%
    \usepackage{parskip}
  }{
    \setlength{\parindent}{0pt}
    \setlength{\parskip}{6pt plus 2pt minus 1pt}}
}{
  \KOMAoptions{parskip=half}}
\makeatother
\usepackage{xcolor}
\makeatletter
\ifx\paragraph\undefined\else
  \let\oldparagraph\paragraph
  \renewcommand{\paragraph}{
    \@ifstar
      \xxxParagraphStar
      \xxxParagraphNoStar
  }
  \newcommand{\xxxParagraphStar}[1]{\oldparagraph*{#1}\mbox{}}
  \newcommand{\xxxParagraphNoStar}[1]{\oldparagraph{#1}\mbox{}}
\fi
\ifx\subparagraph\undefined\else
  \let\oldsubparagraph\subparagraph
  \renewcommand{\subparagraph}{
    \@ifstar
      \xxxSubParagraphStar
      \xxxSubParagraphNoStar
  }
  \newcommand{\xxxSubParagraphStar}[1]{\oldsubparagraph*{#1}\mbox{}}
  \newcommand{\xxxSubParagraphNoStar}[1]{\oldsubparagraph{#1}\mbox{}}
\fi
\makeatother

\providecommand{\tightlist}{%
  \setlength{\itemsep}{0pt}\setlength{\parskip}{0pt}}\usepackage{longtable,booktabs,array}
\usepackage{calc} 
\usepackage{etoolbox}
\makeatletter
\patchcmd\longtable{\par}{\if@noskipsec\mbox{}\fi\par}{}{}
\makeatother
\IfFileExists{footnotehyper.sty}{\usepackage{footnotehyper}}{\usepackage{footnote}}
\makesavenoteenv{longtable}
\usepackage{graphicx}
\usepackage{tikz} 
\usetikzlibrary{positioning, arrows.meta, shapes, calc, fit, decorations.pathmorphing} 
\makeatletter
\def\maxwidth{\ifdim\Gin@nat@width>\linewidth\linewidth\else\Gin@nat@width\fi}
\def\maxheight{\ifdim\Gin@nat@height>\textheight\textheight\else\Gin@nat@height\fi}
\makeatother
\setkeys{Gin}{width=\maxwidth,height=\maxheight,keepaspectratio}
\makeatletter
\def\fps@figure{tbp}
\makeatother

\makeatletter
\@ifpackageloaded{caption}{}{\usepackage{caption}}
\AtBeginDocument{%
\ifdefined\contentsname
  \renewcommand*\contentsname{Table of contents}
\else
  \newcommand\contentsname{Table of contents}
\fi
\ifdefined\listfigurename
  \renewcommand*\listfigurename{List of Figures}
\else
  \newcommand\listfigurename{List of Figures}
\fi
\ifdefined\listtablename
  \renewcommand*\listtablename{List of Tables}
\else
  \newcommand\listtablename{List of Tables}
\fi
\ifdefined\figurename
  \renewcommand*\figurename{Figure}
\else
  \newcommand\figurename{Figure}
\fi
\ifdefined\tablename
  \renewcommand*\tablename{Table}
\else
  \newcommand\tablename{Table}
\fi
}
\@ifpackageloaded{float}{}{\usepackage{float}}
\floatstyle{ruled}
\@ifundefined{c@chapter}{\newfloat{codelisting}{h}{lop}}{\newfloat{codelisting}{h}{lop}[chapter]}
\floatname{codelisting}{Listing}

\makeatother
\makeatletter
\@ifpackageloaded{caption}{}{\usepackage{caption}}
\@ifpackageloaded{subcaption}{}{\usepackage{subcaption}}
\makeatother

\ifLuaTeX
  \usepackage{selnolig}  
\fi
\usepackage[]{natbib}
\usepackage{bookmark}

\IfFileExists{xurl.sty}{\usepackage{xurl}}{} 
\hypersetup{
  pdftitle={Title},
  pdfauthor={Author 1; Author 2},
  pdfkeywords={3 to 6 keywords, that do not appear in the title},
  colorlinks=true,
  linkcolor={blue},
  filecolor={Maroon},
  citecolor={Blue},
  urlcolor={Blue},
  pdfcreator={LaTeX via pandoc}}

\newcommand{\anon}{1}

\date{}
\begin{document}

\def\spacingset#1{\renewcommand{\baselinestretch}%
{#1}\small\normalsize} \spacingset{1}


\if1\anon
{
  \title{\bf Sparsifying Transform Priors in Gaussian Graphical Models}
  \author{Marcus Gehrmann\\
    Department of Mathematical Sciences \\ Norwegian University of Science and Technology\\
    and \\
    Håkon Tjelmeland \\
    Department of Mathematical Sciences \\ Norwegian University of Science and Technology}
  \maketitle
} \fi

\if0\anon
{
  \bigskip
  \bigskip
  \bigskip
  \begin{center}
    {\LARGE\bf Title}
\end{center}
  \medskip
} \fi

\bigskip
\begin{abstract}
Bayesian methods constitute a popular approach for estimating the conditional independence structure in Gaussian graphical models, since they can quantify the uncertainty through the posterior distribution. Inference in this framework is typically carried out with Markov chain Monte Carlo (MCMC). However, the most widely used choice of prior distribution for the precision matrix, the so called G-Wishart distribution, suffers from an intractable normalizing constant, which gives rise to the problem of double intractability in the updating steps of the MCMC algorithm. 

In this article, we propose a new class of prior distributions for the precision matrix, termed ST priors, that allow for the construction of MCMC algorithms that do not suffer from double intractability issues. A realization from an ST prior distribution is obtained by applying a sparsifying transform on a matrix from a distribution with support in the set of all positive definite matrices. We carefully present the theory behind the construction of our proposed class of priors and also perform some numerical experiments, where we apply our methods on a human gene expression dataset. The results suggest that our proposed MCMC algorithm is able to converge and achieve acceptable mixing when applied on the real data.
\end{abstract}

\noindent%
\vfill

\newpage
\spacingset{1.8} 

\section{Introduction}
Gaussian Graphical Models (GGMs) offer a flexible framework for modeling relationships between different continuous variables. They have 
attracted considerable attention in statistical research and have also found their use in a wide range of applications such as genomics \citep{GeneticsApplication}, neural science \citep{NeuralScienceApplication} and power grid analysis \citep{DekaPowerGrid}. For a GGM, we assume that we have data consisting of independent realizations from a mean zero normal distribution that obeys some sort of conditional independence structure induced by an undirected graph. For multivariate Gaussian random vectors, conditional independence between variables implies a zero constraint on the corresponding elements of the precision matrix \citep{rue2005gaussian}. Hence, assuming a sparse graph for the conditional independence structure reduces the number of parameters, which can prevent overfitting and speed up computations.

Typically, both graph and precision matrix are unknown and have to be estimated. The practice of estimating these parameters from data goes back to \citet{Dempster1972}, where he carried out an iterative selection procedure, sequentially adding new edges to the graph. This provides a point estimate of graph and precision. Other procedures for point estimates include backward selection \citep{2000Edwards} and LASSO regularization \citep{MeinhausenLasso}. \citet{Dempster1972} termed the problem of recovering the graph from data \textit{covariance selection}, while the term \textit{structure learning} appears to be more widely used in recent literature \citep{VogelsReviewJASA}. In addition to obtaining point estimates, one is often interested in assessing the uncertainty in the parameters. In a fully Bayesian setup, we assign a joint prior for the graph $G$ and the precision matrix $Q$. This prior can for example be designed in a sequential manner with a marginal prior for the graph and a prior for the precision matrix conditioned on the graph.
Together with the normal likelihood for the data $\mathbf{x}$, we then get a posterior distribution $Q,G|\mathbf{x}$, which can be used to assess parameter uncertainty. Typically, the posterior is not accessible in closed form and we are referred to Markov chain Monte Carlo (MCMC) techniques to infer this distribution.
 Of the two components in the joint prior for $Q$ and $G$, the prior for $Q$: $\pi(Q|G)$ is the most challenging to specify. 
The most popular choice is the so called G-Wishart distribution \citep{Roverato2002}, which has the advantage of being a conjugate prior to the normal likelihood. Nonetheless, inference with this prior has turned out to be difficult, due to lack of an explicit formula for the normalizing constant. 
Multiple different MCMC algorithms for full posterior inference with a G-Wishart prior on the precision matrix have been proposed. One common approach is to use a reversible jump Markov chain Monte Carlo (RJMCMC) algorithm \citep{GreenRJMCMC}, where the acceptance probability contains a ratio of two normalizing constants of the G-Wishart distribution, see e.g. \citet{Dobra_etal2011}. The problem with this framework is that the normalizing constants must be approximated, for instance with a Monte Carlo method \citep{KaysiMassam2005} or a Laplace approximation \citep{LenkoskiDobra2011}. Hence, the error that follows with the approximation of the normalizing constants in the acceptance probability implies that the stationary distribution of the simulated Markov chain deviates from the correct posterior and it is hard to know how large this deviation is.
Exchange type algorithms \citep{MurrayExchange} have enabled MCMC simulation in several other cases with doubly intractable posterior distributions. 
However, applying algorithms of this kind for full posterior inference would require the existence of a direct sampler of the G-Wishart distribution. \citet{LenkoskiDirect2013} proposed an algorithm that was claimed to generate samples from a G-Wishart distribution for arbitrary graphs. Since, many algorithms for full MCMC inference have used this simulation algorithm for the design of MCMC algorithms aiming at inferring the posterior, see for instance \citet{HinneConditionalBayes} and \citet{vanDenBoom2022}. However, the claimed sampler was recently proven to be incorrect \citep{TjelmelandHypothesis}. Hence, the algorithms that relied upon its correctness still lack a direct sampler for valid implementation. As a consequence, full MCMC based Bayesian inference in Gaussian graphical models with the G-Wishart prior remains an unsolved problem.

Recently, \citet{mastrantonio2025newhierarchicaldistributionarbitrary} proposed a new type of prior for the precision matrix, termed the S-Bartlett distribution, with the aim of avoiding the difficulties that arise with the G-Wishart distribution. They construct their prior by specifying the distribution of the free elements of the Cholesky factor of the precision matrix in such a way that the normalizing constant is tractable and then let the non-free elements of the Cholesky factor be specified such that the correct sparsity pattern of the precision matrix is obtained. However, the S-Bartlett distribution has the disadvantage that the prior for the precision matrix conditioned on the graph is dependent on an arbitrary enumeration of the nodes. In particular, this implies that there is no easy way to specify a priori exchangeability between the Gaussian distributed variables.

In the present article, we propose a new class of prior distributions for the precision matrix $Q|G$, that we term \textit{Sparsifying Transform} priors or ST priors for short. Due to their construction, our proposed priors allow for the design of MCMC algorithms that have the correct full posterior $Q,G|\mathbf{x}$ as stationary distribution without any approximations that can distort the limiting distribution of the chain. Moreover, in contrast to the S-Bartlett priors, the ST priors naturally allow for the construction of distributions that do not depend on the enumeration of the nodes and hence may exhibit desired symmetry properties. 

We give general background information in Section \ref{sec:preliminaries}, including an introduction to the G-Wishart distribution. In Section \ref{sec:ST_priors}, we describe the details of the proposed prior distributions. In Section \ref{sec:full_MCMC}, we formulate an MCMC algorithm for full posterior inference with an instance from the proposed distributions as prior for the precision matrix. In Section \ref{sec:results} we outline the results of some numerical experiments and we conclude in Section \ref{sec:concluding_remarks}.
\section{Preliminaries}\label{sec:preliminaries}
The core of a graphical model is the conditional independence graph $G$, that governs the conditional independence between the different variables. We introduce the concept of graphs together with the corresponding notation in Section \ref{subsec:graphs}, while a background on Gaussian graphical models is given in Section \ref{subsec:ggm}. In Section \ref{subsec:priors_for_graph}, we consider some different possible choices for the prior for the graph. We outline the details of the G-Wishart distribution in Section \ref{subsec:G-Wishart} and describe the related Wishart and Inverse Wishart distributions in Section \ref{subsec:wishart_inv_wishart}. In Section \ref{subsec:surjective_map}, we describe the details of a map that is essential for the construction of our proposed class of priors.
\subsection{Graphs and notation}\label{subsec:graphs}
We denote an undirected graph with $G = (V,E)$, where $V = \{1,\ldots,p\}$ is a set of $p$ nodes and $E \subseteq\{(i,j)|i,j\in V, i \neq j\}$ is a set of edges. Whenever $(i,j) \in E$, we say that there is an edge between nodes $i$ and $j$ in the graph. Since we are working with undirected graphs, we can use a symmetry convention in the definition of the edge set such that the statements $(i,j) \in E$ and $(j,i) \in E$ are equivalent. However, when we deal with the size of the edge set $|E|$, we count $(i,j)$ and $(j,i)$ as one edge. When we work with precision matrices related to Gaussian graphical models, it is convenient to also make use of an extended edge set $\mathcal{V}$, that in addition to the edges contains all pairs on the form $(i,i)$, for all $i \in V$. That is
\begin{equation}\label{eq:extended_edge_set}
    \mathcal{V}\stackrel{\Delta}{=} E\cup\left\{(i,i)| i\in V\right\}.
\end{equation}
If we have a (possibly stochastic) $p$-dimensional vector $x \in \mathbb{R}^{p}$ that is indexed over the set of nodes $V$, we denote with $x_i$ the component of $x$ belonging to node $i$. Furthermore, for arbitrary $A \subseteq V$, we denote with $x_{A}$ the restriction of $x$ onto $A$ and with $x_{-A}$ the restriction of $x$ onto $V\setminus A$. 
More formally, we define
\begin{equation*}
    x_A = [x_i|i \in A]\hspace{5mm}\text{and}\hspace{5mm}x_{-A} = [x_i|i \notin A].
\end{equation*}
Likewise, for matrix $P \in \mathbb{R}^{p\times p}$, we denote with $P_A$ the submatrix that we get by extracting rows and columns according to the set $A \subseteq V$ and with $P_{A,B}$ the submatrix we get by extracting rows according to $A$ and columns according to $B \subseteq V$. Formally,
\begin{equation*}
   P_A = \left[P_{ij};i,j\in A\right] \hspace{5mm}\text{and}\hspace{5mm}
   P_{A,B} = \left[P_{ij};i \in A, j \in B\right].
\end{equation*}
A \textit{clique} $\mathcal{C}\subseteq V$ is a set of nodes such that there is an edge between all distinct pairs of nodes in the clique. 

Let $p$ be an arbitrary positive integer. We denote with $\mathbb{P}$ the set of all positive definite matrices of size $p$. 
Since the dimension $p$ of the matrix is arbitrary or implicit, it is omitted from the notation.
For a graph $G$, we denote with $\mathbb{P}(G)$ the set of positive definite matrices with a zero constraint on the elements corresponding to the elements not belonging to the extended edge set $\mathcal{V}$. That is 
\begin{equation*}
    \mathbb{P}(G) = \left\{Q \in \mathbb{P} | Q_{ij} = 0 \,\,\,\forall (i,j) \notin \mathcal{V}\right\}.
\end{equation*}
\subsection{Gaussian graphical models}\label{subsec:ggm}
In a Gaussian graphical model, we assume to have a graph $G = (V,E)$ with which we associate a $p$-dimensional stochastic variable $x$, where the elements are indexed over the set of nodes $V$. Since the model is Gaussian, $x$ follows a Gaussian distribution. It is custom in the field to assume this distribution to be mean zero although including a non-zero mean $\mu$ into the model is possible in principle. In addition to being Gaussian, $x$ obeys a conditional independence property that is governed by the structure of the graph.
More precisely, if $(i,j) \notin E$, then $x_i$ is conditionally independent of $x_j$ given all other variables. That is
\begin{equation}\label{eq:pairwise_independence}
    (i,j) \notin E \implies x_i \perp x_j \mid x_{-\{i,j\}}.
\end{equation}
 The pairwise conditional independence feature of \eqref{eq:pairwise_independence} is equivalent to the corresponding element in the precision matrix $Q$ being zero: $Q_{ij} = 0$. Hence, the lack of edges in the graph obeys a one-to-one correspondence with the zero structure of the precision matrix. Stated explicitly,
 \begin{equation*}
     (i,j) \notin \mathcal{V} \implies Q_{ij} = 0.
 \end{equation*}
 Thus, a precision matrix $Q$ corresponding to a GGM with graph $G$ fulfills $Q \in \mathbb{P}(G)$.
 When applying GGMs in practice, we assume to have data 
\begin{equation}\label{eq:X_matrix}
\mathbf{x} = \begin{bmatrix}
  x^{(1)} & x^{(2)} & \ldots & x^{(m)}  
\end{bmatrix}^T,
\end{equation}
which is $m \times p$ where $m$ is the number of observations and the rows of $\mathbf{x}$ are independent realizations of a stochastic variable $x \sim \mathcal{N}(0,Q^{-1})$. The goal in structure learning is to recover the conditional independence graph $G$, and most often also the precision matrix $Q$, based on the data $\mathbf{x}$. 

As briefly outlined in the introduction, the Bayesian approach for structure learning requires a joint prior on the conditional independence graph and the precision matrix. Due to the correspondence between the graph and the zero constraints on the precision matrix, the prior is naturally constructed in a sequential manner as
\begin{equation}\label{eq:sequential_prior}
    \pi(G,Q) = \pi(Q|G)\pi(G).
\end{equation}
As a consequence of the discrete nature of the graph, there are many viable options for the marginal prior $\pi(G)$. We consider a few in Section \ref{subsec:priors_for_graph}. To specify a prior for the precision matrix $Q|G$ comes with greater difficulties. A large part of the difficulty stems from the fact that this prior distribution must have its support in $\mathbb{P}(G)$, and this can be regarded as a non-trivial domain. The most widely used prior for the precision matrix is the G-Wishart distribution \citep{Roverato2002}. We give a brief introduction to this distribution in Section \ref{subsec:G-Wishart} below.
\subsection{Prior distributions for the graph}\label{subsec:priors_for_graph}
A common choice for the prior for $G$ is to assign equal probabilities to all possible graphs, see for instance \citet{WangLiEfficient}. Another option is to assign independent Bernoulli priors for the presence of edges between any pair of nodes in the graph \citep{VogelsReviewJASA}. If the probability for edge inclusion is set to a half, we get the uniform case described above.
A problem with the independent Bernoulli priors for the presence of edges is that it is rather informative with regards to the total number of edges in the graph. The total number of edges follows a binomial distribution, which tends to have a lot of the probability mass centered around the mean if the number of nodes is large enough. Another possibility is therefore to use a so called double uniform prior for the graph. This means that we assume a uniform prior for the number of edges, $\pi(|E|) \propto 1$, while the distribution for the graph conditioned on the number of edges is uniform among all possible choices. That is, the probability mass function $\pi(G)$ is given by 
 \begin{equation*}
     \pi(G) = \frac{1}{E_{\text{max}} + 1}\cdot\frac{1}{\binom{E_{\text{max}}}{|E|}},
 \end{equation*}
with $E_{\text{max}} = \frac{|V|(|V| - 1)}{2}$. As far as we know, this choice of prior for $G$ has not been used in connection with structure learning before, but similar constructions have appeared in other contexts, such as in \citet{ChipmanCart} for Bayesian CART models and in \citet{LuoMarkovMesh} for neighborhood structures in Markov mesh models.

The idea behind the double uniform prior can clearly be generalized by assigning a non-uniform prior to the number of edges, $|E| \sim \pi(|E|)$, while retaining the uniform prior for the graph conditioned on the number of edges. One possibility that favors sparsity by penalizing many edges is to let $\pi(|E|) \propto \theta^{|E|}$ for $\theta \in (0,1)$. This choice of prior will be referred to as a truncated geometric prior in the following.
\subsection{The G-Wishart distribution}\label{subsec:G-Wishart}
We say that $Q|G$ is G-Wishart distributed with parameters $\delta$ and $D$ or $Q|G \sim \mathcal{W}_{G}(\delta,D)$ if it has density
\begin{equation}\label{eq:G_Wishart_density}
    \pi(Q|G) = \frac{1}{I_G(\delta,D)}|Q|^{\frac{\delta - 2}{2}}e^{-\frac{1}{2}(Q,D)}\cdot\mathbb{I}\left(Q \in \mathbb{P}(G)\right)
\end{equation}
with $\delta > 2$ and $D \in \mathbb{P}$, while $I_G(\delta,D)$ is a normalizing constant and where $(A,B) = \text{tr}(A^TB)$ denotes the matrix inner product. 
One should note that since $Q$ is restricted to be symmetric and to have $Q_{ij} = 0$ whenever $(i,j) \notin \mathcal{V}$, the number of free parameters in $Q$ is $p + |E|$. The expression in \eqref{eq:G_Wishart_density} should be understood as a density for these free elements. This notation is standard practice for G-Wishart distributions and in the following we use the same convention whenever treating densities with support in $\mathbb{P}$ or $\mathbb{P}(G)$.
The motivation behind choosing the G-Wishart prior is mainly that it is conjugate to the normal likelihood. If $Q|G \sim \mathcal{W}_{G}(\delta,D)$ and $x^{(1)},x^{(2)},\ldots,x^{(m)} \stackrel{\text{i.i.d}}{\sim} \mathcal{N}(0,Q^{-1})$, then
\begin{equation*}
    Q|G,\mathbf{x} \sim \mathcal{W}_G(\delta + m,D + \mathbf{x}^T\mathbf{x}),
\end{equation*}
where the matrix $\mathbf{x}$ is structured as described in \eqref{eq:X_matrix}.
However, the normalizing constant of the G-Wishart distribution $I_G(\delta,D)$ poses a major challenge for the use of the G-Wishart distribution in Bayesian inference. Traditionally, the normalizing constant has been approximated. \citet{UhlerLenkoskiNormalizing} provided a formula for $I_G(\delta,D)$ in the general case. However, the formula contains nested infinite sums which means that it only can be efficiently computed for certain types of graphs and values of the matrix parameter $D$ and as far as we know, there are no examples where the exact formula has been implemented for inference in practice. More recently, \citet{wong2025newwayevaluategwishart} extended the class of combinations of graphs and hyperparameter $D$ for which the normalizing constant can be computed by using Fourier-based methods. Yet, a computationally viable method to calculate the normalizing constant in the general case does not exist as of today.

In addition to the attempts at viable methods to compute the normalizing constant, there has been some work on algorithms for obtaining samples from the G-Wishart distribution. \citet{WangCarvalhoHypInvWSim} proposed a rejection sampler for direct sampling, but it has turned out less useful due to low acceptance rate even for medium sized matrix dimensions \citep{Dobra_etal2011}. One later contribution is the claimed direct sampler by \citet{LenkoskiDirect2013}, which \citet{TjelmelandHypothesis} proved incorrect.

When employing a G-Wishart distribution in the wider context of MCMC inference in GGMs, we typically want to compute the ratio of posterior densities in two points, say $(Q',G')$ and $(Q,G)$, and in the case of the G-Wishart distribution where the normalizing constant of $\pi(Q|G)$ is intractable and varies with the graph, this ratio cannot be computed exactly. This poses a challenge to the use of the G-Wishart distribution as a prior for the precision matrix in this context.
\subsection{The Wishart and Inverse Wishart distributions}\label{subsec:wishart_inv_wishart}
If the graph $G$ is full and we replace $Q$ with $S$, the density in \eqref{eq:G_Wishart_density} transforms into
\begin{equation}\label{eq:Wishart_density}
    \pi(S) = \frac{1}{I_{p}(\delta,D)}|S|^{\frac{\delta - 2}{2}}e^{-\frac{1}{2}(S,D)}\cdot\mathbb{I}\left(S \in \mathbb{P}\right),
\end{equation}
where the normalizing constant $I_{p}(\delta,D)$ depends on the dimension $p$ only.
The distribution associated with this density constitutes a special case of the G-Wishart distribution, which is called the Wishart distribution, and is denoted with $\mathcal{W}_{p}(\delta,D)$. For the Wishart distribution, the normalizing constant $I_p(\delta,D)$ has a closed form expression and can hence be efficiently computed. Yet another related distribution is the Inverse Wishart distribution. We say that if $S \sim \mathcal{W}_p(\delta,D)$, then its inverse $T = S^{-1}$ follows an Inverse Wishart distribution of size $p$ with parameters $\delta$ and $D$ or $T \sim \mathcal{IW}_{p}(\delta,D)$. The corresponding density for this distribution becomes
\begin{equation}\label{eq:invWishart_density}
    \pi(T) = 
    \frac{1}{I_p(\delta,D)}
|T|^{-\frac{\delta + 2p}{2}}e^{-\frac{1}{2}(T^{-1},D)}
\mathbb{I}(T \in \mathbb{P}).
\end{equation}
With this parametrization of the distribution and provided that $\delta > 2$, the expected value for $T \sim \mathcal{IW}_{p}(\delta,D)$ is given by
\begin{equation}\label{eq:expected_value_invwish}
    E[T] = \frac{D}{\delta - 2},
\end{equation}
while the standard deviations for the diagonal elements are given by
\begin{equation}\label{eq:sd_invwish}
    \text{SD}[T_{ii}] = \sqrt{\frac{2}{\delta - 4}}\cdot\frac{D_{ii}}{\delta - 2},
\end{equation}
provided that $\delta > 4$ \citep[Chapter 5]{PressMultivariateAnalysis}.
\subsection{A surjective map}\label{subsec:surjective_map}
This section outlines the details of a surjective map between two matrix related subspaces that plays an essential role in the construction of our prior distribution. This map, termed positive definite completion or PD-completion for short, has occurred frequently before in connection with the G-Wishart distribution, for instance in \citet{LenkoskiDirect2013}. It was also discussed by \citet{Roverato2002}. 
The goal is to construct a function that, for an arbitrary graph $G$, constitutes a map from the set of positive definite matrices $\mathbb{P}$ to the set of positive definite matrices with a zero structure induced by the graph $G$. Since the map depends on the graph $G$, we denote it with $\text{PD}_{G}(\cdot)$ and we have that
\begin{equation}\label{eq:pd_map_uniqueness}
    \text{PD}_G: \mathbb{P} \to \mathbb{P}(G).
\end{equation}
The nature of the map in \eqref{eq:pd_map_uniqueness} is related to matrix inversion. For arbitrary $\Sigma \in \mathbb{P}$, the inverse $\Sigma^{-1}$ is also positive definite, but does not necessarily have a sparsity pattern according to the graph $G$. Thus, there is in most cases not a $Q \in \mathbb{P}(G)$ such that $Q^{-1} = \Sigma$. However, if we relax the requirement that $Q^{-1}$ should be equal to $\Sigma$ and only demand that $Q^{-1}$ should be equal to $\Sigma$ at the indices corresponding to the extended edge set $\mathcal{V}$, it turns out that there is one and only one $Q \in \mathbb{P}(G)$ that fulfills this requirement. This statement is formalized in Theorem \ref{thm:surjective_map}. For a proof and more details, see \citet{Grone1984}.
\begin{theorem}\citep{Grone1984}\label{thm:surjective_map}
    Let $G$ be a graph and $\mathcal{V}$ be the corresponding extended edge set. For any $\Sigma \in \mathbb{P}$, there is one and only one $Q \in \mathbb{P}(G)$ such that $\Sigma_{ij} = (Q^{-1})_{ij} \,\,\forall (i,j) \in \mathcal{V}$.
\end{theorem}
\noindent
Since there is one and only one $Q$ that fulfills the property in Theorem \ref{thm:surjective_map}, we can define a function that for each $\Sigma \in \mathbb{P}$ outputs the corresponding $Q \in \mathbb{P}(G)$. This is our definition of $\text{PD}_G(\cdot)$ and this is formalized in Definition \ref{def:pd_completion_map}.
\begin{definition}\label{def:pd_completion_map}
    Let $\Sigma \in \mathbb{P}$. For any graph $G$, $\text{PD}_{G}(\Sigma)$ denotes the unique $Q$ that fulfills the requirements specified in Theorem \ref{thm:surjective_map}.
\end{definition}
\noindent
Since the dimension of $\mathbb{P}(G)$ is smaller than the dimension of $\mathbb{P}$, the mapping is many-to-one. Furthermore, if we let $Q \in \mathbb{P}(G)$ be arbitrary and define $\Sigma = Q^{-1}$, then $\text{PD}_G(\Sigma) = Q$. This concludes the surjectivity.
Let us now assume that we have two matrices $\Sigma$ and $\Sigma'$, both of them in $\mathbb{P}$, such that they coincide in all of $\mathcal{V}$. That is, $\Sigma_{ij} = \Sigma'_{ij}\,\,\forall (i,j) \in \mathcal{V}$. Then, $\text{PD}_G(\Sigma) = \text{PD}_G(\Sigma')$. Therefore, the value of $\text{PD}_G(\Sigma)$ only depends on the elements of $\Sigma$ that correspond to indices in $\mathcal{V}$. The elements of $\Sigma$ corresponding to the complement of $\mathcal{V}$ are redundant.

Theorem \ref{thm:surjective_map} only guarantees the existence of the map $\text{PD}_G(\cdot)$, but does not say anything about how to compute it and we lack an analytical expression for the function. Instead, we have to rely on iterative algorithms. Two main algorithms have been considered in the literature, the Iterative Proportional Scaling (IPS) algorithm \citep{LauritzenGraphicalModels1996} that operates on submatrices of $Q$ corresponding to cliques of $G$ and the algorithm proposed by \citet{hastie2009elements}, that was subsequently deployed in the algorithm of \citet{LenkoskiDirect2013}. The latter operates on the columns of $W = Q^{-1}$ and has been the main choice in the recent literature and we use it for the numerical experiments in this article. In the following, we refer to this algorithm as the Hastie algorithm.
\section{ST priors}\label{sec:ST_priors}
In the present section, we describe the general idea behind our new class of prior distributions, that we term ST priors, standing for Sparsifying Transform priors. This class of distributions is defined in Section \ref{subsec:st_prior_def}, and in Section \ref{subsec:using_st_priors} we outline how the nature of this class of distributions can be exploited in full Bayesian inference for GGMs when using these distributions as priors. The reason for terming this class of distributions ST priors is that we obtain a realization from the distribution by applying the transform defined in Definition \ref{def:pd_completion_map} to a realization from an arbitrary distribution with support in $\mathbb{P}$ and by the definition of this transform, it obtains sparsity while taking a full positive definite matrix as input. 
\subsection{Definition of the class of distributions}\label{subsec:st_prior_def}
We give a general definition of the class of ST priors in Definition \ref{def:st_prior_dist}.
\begin{definition}\label{def:st_prior_dist}
    Let $G = (V,E)$ be an undirected graph. Moreover, let $\tilde{\pi}(\cdot)$ be an arbitrary distribution with support in $\mathbb{P}$. If $\Sigma \sim  \tilde{\pi}(\Sigma)$ and Q = $\text{PD}_{G}(\Sigma)$, then we say that Q is ST distributed according to graph $G$ and distribution $\tilde{\pi}$ or
    \begin{equation*}
        Q \sim ST(G;\tilde{\pi}).
    \end{equation*}
\end{definition}
\noindent
Regarding the choice of $\tilde{\pi}$, we are offered a lot of flexibility as long as the restriction of support in $\mathbb{P}$ is fulfilled. Natural choices include the Wishart and Inverse Wishart distributions described in \eqref{eq:Wishart_density} and \eqref{eq:invWishart_density} respectively. 
According to \citet{LenkoskiDirect2013}, an Inverse Wishart distribution for $\tilde{\pi}(\cdot)$ yields a G-Wishart distribution for arbitrary $G$, but this claim was refuted in \citet{TjelmelandHypothesis}.
The present article mainly focuses on the case of $\tilde{\pi}(\cdot)$ being a Wishart distribution.

Due to the properties of the transform $\text{PD}_{G}(\cdot)$, the ST distributions have their support in $\mathbb{P}(G)$, which makes them valid priors for precision matrices in a GGM.
Furthermore, a consequence of Theorem \ref{thm:surjective_map} in the context of a GGM is that there is a one-to-one correspondence between the prior distribution for the precision matrix $Q$ and the prior distribution for the elements of $Q^{-1}$ corresponding to $\mathcal{V}$. Hence, if we fix the graph $G$, essentially any distribution with support in $\mathbb{P}(G)$, say $f$, can in theory be represented within the ST prior framework, by specifying the marginal of $\tilde{\pi}$ at $\mathcal{V}$ in correspondence with the chosen $f$ and then specify some conditional distribution for the remaining elements of $\Sigma$, conditioned on the elements in $\mathcal{V}$, such that the support in $\mathbb{P}$ is obtained. In practice however, this is difficult due to the intricacy of the PD map. Moreover, as will be outlined in Section \ref{subsec:using_st_priors}, we will assume that the distribution $\tilde{\pi}$ is independent of $G$ in order to facilitate inference. This imposes further constraints.
 
Note that for an ST prior distribution $ST(G;\tilde{\pi})$, it is non-trivial to obtain an expression for the density of $Q|G$.
Nonetheless, the nature of this class of distributions still allows us to use them for full MCMC inference in GGMs, without the need to evaluate their density. The details are outlined in Sections \ref{subsec:using_st_priors} and \ref{sec:full_MCMC} below.
 \subsection{Using ST priors in Bayesian structure learning}\label{subsec:using_st_priors}
 We can now outline how we can employ the ST priors defined in Section \ref{subsec:st_prior_def} in a joint prior for $G$ and $Q$: $\pi(Q,G)$ and how a clever use of auxiliary variables within this framework can be exploited when performing inference.

 We employ the ordinary sequential framework for the joint prior for graph and precision matrix described in \eqref{eq:sequential_prior}. The prior for the graph $G$ can be arbitrary, while we use a prior from the class of distributions described in Section \ref{subsec:st_prior_def}. That is
 \begin{equation}\label{eq:prior_Q_conditioned_on_G}
     Q|G \sim ST(G;\tilde{\pi})
 \end{equation}
 for suitable choice of $\tilde{\pi}$.
 Note that $\tilde{\pi}$ in \eqref{eq:prior_Q_conditioned_on_G} is not a function of the graph $G$.
 Together with the normal likelihood for the data $\mathbf{x}|Q \sim \mathcal{N}(0,Q^{-1})$, we wish to infer the posterior
\begin{equation*}
    \pi(Q,G|\mathbf{x}) \propto \pi(G)\pi(Q|G)\pi(\mathbf{x}|Q).
\end{equation*}
To do this, we describe the model in a different fashion with the help of auxiliary variables.
In this alternative formulation of the model, the prior for the graph $\pi(G)$ remains intact. In addition, we have a parameter $\Sigma \in \mathbb{P}$, that hence is a full positive definite matrix. We let the prior for $\Sigma$ be $\tilde{\pi}(\cdot)$. That is,
\begin{equation*}
    \Sigma \sim \tilde{\pi}(\Sigma),
\end{equation*}
where $\tilde{\pi}$ is the distribution that defines the prior for $Q$ in the first model formulation in \eqref{eq:prior_Q_conditioned_on_G}. We furthermore let $G$ and $\Sigma$ be a priori independent
 \begin{equation}\label{eq:joint_G_Sigma_prior}
     \pi(\Sigma,G) = \tilde{\pi}(\Sigma)\cdot\pi(G).
 \end{equation}
As before the observations are mean zero normal, $\mathbf{x} \sim \mathcal{N}(0,Q^{-1})$. In this formulation, we let  \begin{equation}\label{eq:Q_as_function_of_Sigma}
 Q = \text{PD}_G(\Sigma). 
 \end{equation}
 Using Definition \ref{def:st_prior_dist}, we can see that with this alternative formulation, the prior for $Q|G$ is an ST distribution with $\tilde{\pi}$ as distribution parameter. That is, \eqref{eq:prior_Q_conditioned_on_G} holds. Since both the likelihood and the prior for $G$ remain the same, this alternative formulation of the model is equivalent to the original one. Note that in the original model formulation, the parameters are $G$ and $Q$. In the alternative formulation, the parameters are $G$ and $\Sigma$, where $Q$ that appears in the likelihood   is a function of the parameters using \eqref{eq:Q_as_function_of_Sigma}. Since the likelihood depends on $Q$ only, the distributions $\mathbf{x}|\Sigma,G$ as well as the corresponding posterior $\Sigma,G|\mathbf{x}$ are well defined.
The two formulations of the model are a priori equivalent.
Furthermore, posterior inference in the alternative model formulation can be exploited for inference in the original model formulation.
If we can obtain a sample
\begin{equation*}
    \Sigma^{*},G^{*} \sim \pi(\Sigma,G|\mathbf{x}),
\end{equation*}
then
\begin{equation}\label{eq:correct_posterior_for_Q_G}
    Q^{*},G^{*} = \text{PD}_{G^{*}}(\Sigma^{*}),G^{*} \sim \pi(Q,G|\mathbf{x}).
\end{equation}
That \eqref{eq:correct_posterior_for_Q_G} is valid follows from the fact that we always can interchange the order of transformation and conditioning without affecting the distribution. 
In this particular case, the transformation is given by a combination of the PD-completion applied on $\Sigma$ and an identity map for the graph. Thus, if we have a method to sample from $\Sigma,G|\mathbf{x}$, \eqref{eq:correct_posterior_for_Q_G} yields a way of sampling from $Q,G|\mathbf{x}$. 

The reason behind inference in the alternative model formulation with $\Sigma$ and $G$ being favorable comes from the a priori independence between the parameters stated in \eqref{eq:joint_G_Sigma_prior}. In MCMC, we typically need to compute a ratio of posterior densities in two different points, in order to compute an acceptance probability. 
 In the case of the G-Wishart distribution this ratio cannot be computed, without knowing the normalizing constants. When $\Sigma$ and $G$ are a priori independent, this problem does no longer exist. Within this framework, the ratio between posterior densities in $(\Sigma',G')$ and $(\Sigma,G)$ can be written as
\begin{equation*}
    \frac{\pi(G')\tilde{\pi}(\Sigma')\pi\left(\mathbf{x}|G',\Sigma'\right)}{\pi(G)\tilde{\pi}(\Sigma)\pi\left(\mathbf{x}|G,\Sigma\right)}
\end{equation*}
and we can see that this ratio can be computed even if we do not know the normalizing constant for the prior for $\Sigma$.

However, the use of ST priors comes with a price. Since our ST prior is not conjugate to the normal likelihood it is more difficult to design proposal distributions in an MCMC setting that are informed by the data. In addition, the fact that we are using auxiliary variables means that for most $G$, there are elements of $\Sigma$ that do not affect the observations. This can for instance cause problems due to higher posterior variance for some elements of $\Sigma$ than for others and a slow exploration of the posterior when using MCMC.

\section{MCMC inference with ST priors}\label{sec:full_MCMC}
We assume to have a prior for the graph $G$ and that the conditional prior for $Q|G$ is given by an ST prior distribution treated in Section \ref{sec:ST_priors}. When performing posterior inference within this framework, we employ the ideas outlined in Section \ref{subsec:using_st_priors}, with a joint prior for parameters $\Sigma \in \mathbb{P}$ and $G$, such that the ST prior for $Q|G$ is implicit. We then simulate a Markov chain with the posterior $\Sigma,G|\mathbf{x}$ as stationary distribution. Starting in $(\Sigma^{(1)},G^{(1)})$, we let the chain run for $s$ iterations such that we obtain samples
\begin{equation*}
    (\Sigma^{(1)},G^{(1)}),\ldots,(\Sigma^{(s)},G^{(s)}).
\end{equation*}
By applying the transform in \eqref{eq:correct_posterior_for_Q_G}, we get a corresponding set of samples
\begin{equation*}
    (Q^{(1)},G^{(1)}),\ldots,(Q^{(s)},G^{(s)}).
\end{equation*}
If the Markov chain converges fast enough, after discarding a number of initial samples corresponding to a burn-in, the remaining $(Q^{(i)},G^{(i)})$ will be (approximate) samples from the posterior $Q,G|\mathbf{x}$.

When proposing new states in the chain, we alternate between two types of proposals. Either we propose a new $\Sigma$, independently of the current state of the graph, while keeping the graph unchanged, i.e. we propose $\Sigma^{*}$ from a proposal distribution $q(\Sigma^{*}|\Sigma)$. Alternatively, we propose a new graph, independently of the current state of $\Sigma$, while keeping $\Sigma$ unchanged, i.e. we propose $G^{*}$ from a proposal distribution $q(G^{*}|G)$. Both of these updates can be seen as instances of the Metropolis-Hastings algorithm \citep{Metropolis1953,Hastings1970}, where the proposed new state is either accepted or rejected with a probability $\alpha$. Standard theory can be applied to compute the acceptance probability. It should be noted that other, more complicated, proposal distributions could be applied. One could for instance propose joint changes in both $\Sigma$ and $G$. Here, we focus on these two simple proposals, where we update one variable at the time.
Proposing an update of the graph is typically done by proposing to add or remove an edge with a suitable transition kernel. While proposing a new graph is simple in principle, a new graph also implies a new precision matrix $Q$, which means that we have to recompute $Q$ via PD-completion once again in order to reevaluate the likelihood in the acceptance probability. 
\subsection{Updating $\Sigma$}\label{subsec:update_Sigma}
For the proposal distribution for $\Sigma$, $q(\Sigma^{*}|\Sigma)$, we can either propose changes in all elements of $\Sigma$ or propose changes in smaller blocks. 
When the size of the graph $p$ is large, such that the parameter space for $\Sigma$ is high-dimensional, proposing changes in smaller blocks is preferable to avoid a high rejection rate. 
We describe a scheme for such block proposals below.
If the size of a block is equal to $p$, we get a proposed update of all of $\Sigma$ as a special case.

To ensure the positive definiteness of the proposed new $\Sigma$, the proposed change is done in the domain of the Schur complement for the part of $\Sigma$ that we wish to update.
More formally, let $B \subseteq V$ be an arbitrary subset of the nodes. We permute the rows and columns of $\Sigma$ such that $\Sigma_{B}$ ends up in the upper left corner. We then get the block decomposition
\begin{equation*}
    \Sigma = \begin{bmatrix}
        \Sigma_B & \Sigma_{B,V\setminus B} \\
        \Sigma_{V\setminus B,B} & \Sigma_{V\setminus B}
    \end{bmatrix}.
\end{equation*}
For $\Sigma_{B,V\setminus B}$ and $\Sigma_{V\setminus B} \in \mathbb{P}$ fixed, the positive definiteness of $\Sigma$ is equivalent to the positive definiteness of the Schur complement
\begin{equation*}
    S_{B} \stackrel{\Delta}{=} \Sigma_B - \Sigma_{B,V\setminus B}\Sigma_{V\setminus B}^{-1}\Sigma_{V\setminus B,B}.
\end{equation*}
We can exploit this fact to use an Inverse Wishart distribution to update the Schur complement corresponding to $B$. This yields an indirect update of the block $\Sigma_B$ that maintains the positive definiteness of $\Sigma$ as a whole. More precisely, we propose a new Schur complement corresponding to the subset $B$ from a proposal distribution $q(S_{B}^{*}|S_B)$.
We want $q(S_B^{*}|S_B)$ to fulfill two properties. Firstly, inspired by random walk proposals, we want the expected value of the proposal to coincide with the current value, such that $E[S_B^{*}|S_B] = S_B$. Secondly, we want
\begin{equation*}
    \frac{\text{SD}[(S_B^{*})_{ii}|S_B]}{(S_B)_{ii}} = c \hspace{1cm}\forall i,
\end{equation*}
where $c$ is a tuning parameter of our choice. That is, we want the proposal to be centered at the current parameter value, while being able to regulate the proposal standard deviations of the diagonal terms as a fraction of the present values. By using \eqref{eq:expected_value_invwish} and \eqref{eq:sd_invwish}, we can see that choosing
\begin{equation*}
    S_B^{*}|S_B \sim \mathcal{IW}_{|B|}(k + 2,k\cdot S_B)
\end{equation*}
with $k = \frac{2}{c^2} + 2$ satisfies the two desired properties of the proposal distribution.
The proposed new value of $\Sigma$: $\Sigma^{*}$ is obtained as
\begin{equation*}
    \Sigma^{*} = 
    \begin{bmatrix}
        S_B^{*} + \Sigma_{B,V\setminus B}\Sigma_{V\setminus B}^{-1}\Sigma_{V\setminus B,B} & \Sigma_{B,V\setminus B} \\
        \Sigma_{V\setminus B,B} & \Sigma_{V\setminus B}
    \end{bmatrix}.
\end{equation*}
When choosing which blocks to update, we can select a set of blocks in advance $B_1,\ldots,B_l$ such that each distinct pair of nodes in $V$ occurs in at least one of the blocks and then propose updates for these blocks sequentially in a predefined deterministic order. Alternatively, we can select blocks $B$ with $|B| > 1$ randomly in each updating step. Both approaches lead to Markov chains that are irreducible with respect to $\Sigma$.
\section{Results}\label{sec:results}
In order to evaluate our proposed prior with the associated inference procedure in practice, we apply it to a real dataset. We use the gene expression data set used by \citet{MohammadiBayesStructLearning} and \citet{vanDenBoom2022}, that was originally described by \citet{StrangerGeneExpression}. The goal is to show that our proposed algorithm converges and mixes acceptably, while also comparing the results for different choices of priors for the graph. We will therefore apply our method with four different choices of prior distributions for $G$, namely a double uniform prior, a uniform prior and two truncated geometric priors with $\theta = 0.9901$ and $0.9804$ respectively. The values of $\theta$ for the truncated geometric priors are chosen so that the expected number of edges are $100$ and $50$ respectively.
In the following, we refer to the algorithm proposed in the present article as STMH, standing for Sparsifying Transform Metropolis-Hastings.

For comparison, we apply the same dataset to two other algorithms that perform posterior inference with the G-Wishart prior, namely the WWA algorithm described by \citet{vanDenBoom2022} as well as the standard algorithm available through the \textit{BDgraph} package \citep{BDGraph}. Sections \ref{subsec:data+normalization} to \ref{subsec:results_STMH} are devoted to the numerical experiments with the STMH algorithm, while the results from the WWA and \textit{BDgraph} algorithms are presented in Section \ref{subsec:res_wwa+BDgraph}.
\subsection{Details about the data and normalization}\label{subsec:data+normalization}
A subset of the gene expression dataset corresponding to the $p = 100$ most variable genes is accessible through the \textit{BDgraph} package and we collect the data from there. We can look at even smaller subsets of the data by selecting the in turn most variable genes from the 100 available variables. In this article we decide to constrain ourselves to the case of $p = 50$. In order to obtain data that marginally follows a standard Gaussian distribution, we process the raw data with the quantile normalization method used by \citet{vanDenBoom2022}. 
\subsection{Prior for $Q|G$}
The standard choice of hyperparameters for the G-Wishart prior is letting $\delta = 3$ and $D = I$ \citep{VogelsReviewJASA}. That is,
\begin{equation}\label{eq:G_wishart_prior_real_data}
    Q|G \sim \mathcal{W}_{G}(3,I_{50}).
\end{equation}
The most natural analogue of this choice in our setting with ST priors would be to let $\tilde{\pi}(\Sigma) \sim \mathcal{IW}_{50}(3,I_{50})$. If the claimed sampler by \citet{LenkoskiDirect2013} were correct, this choice of $\tilde{\pi}$ would in fact yield the exact same prior as \eqref{eq:G_wishart_prior_real_data}. However, numerical experiments on simulated data suggest that the choice of an Inverse Wishart distribution for $\tilde{\pi}$ can yield a multimodality in the posterior for $\Sigma$, especially in the elements with low posterior edge probability. To avoid this complication, we instead adopt a Wishart prior for $\Sigma$ with $\delta = 1$ and $D = 50\cdot I_{50}$, such that $Q|G \sim ST(G;\tilde{\pi})$ with
\begin{equation*}
    \tilde{\pi}(\Sigma) \sim \mathcal{W}(1,50\cdot I_{50}).
\end{equation*}
Regardless of prior for the graph, the prior for the precision matrix is the same.
\subsection{Implementation and tuning parameters}
We start with an empty graph and initialize $\Sigma$ at the identity matrix $I_{50}$.
We alternate proposed updates of the graph with proposed updates of blocks of $\Sigma$. One iteration of the algorithm is defined as one proposed update of the graph and one round of block proposals for $\Sigma$.
We propose updates of the graph by either proposing to add an edge or remove an edge. If the graph is neither full nor empty, whether to add or remove an edge are assigned equal probabilities. If the graph is empty, we propose to add an edge with probability one, whereas we propose to remove an edge with probability one in the case of having a full graph. Which edge to add or remove is drawn uniformly among all possible choices. That is, the proposal probability for going from graph $G = (V,E)$ to graph $G^{*} = (V,E^{*})$, denoted $q(G^{*}|G)$, is given by
\begin{equation*}
\begin{split}
    q(G^{*}|G) = \mathbb{I}(|E^{*}| - |E| = 1, E\subset E^{*})\frac{1}{2 - \mathbb{I}(|E| = 0)}\cdot\frac{1}{E_{\text{max}} - |E|}\\
    + \mathbb{I}(|E^{*}| - |E| = -1,E^{*}\subset E)\frac{1}{2 - \mathbb{I}(|E| = E_{\text{max}})}\cdot\frac{1}{|E|}.
    \end{split}
\end{equation*}
When proposing changes in $\Sigma$, we make use of the block update outlined in Section \ref{subsec:update_Sigma}. In each iteration, we propose seven updates of $\Sigma$ in randomly selected blocks of size $20$. The tuning parameter $c$ is set to $1/35$.
This choice of $c$ was decided through tuning of the acceptance rate to a value between $0.2$ and $0.3$ \citep{RobertsRosenthalOptimalScaling}.
The PD-completion step of the algorithm is carried out with the Hastie algorithm.

The STMH algorithm is run for $1\,000\,000$ iterations with $100\,000$ iterations considered as burn-in, regardless which of the four possible priors for the graph we use.
\subsection{Results for STMH}\label{subsec:results_STMH}
\begin{figure}
    \centering
    \begin{subfigure}{0.45\linewidth}
        \centering
        \includegraphics[width=\linewidth]{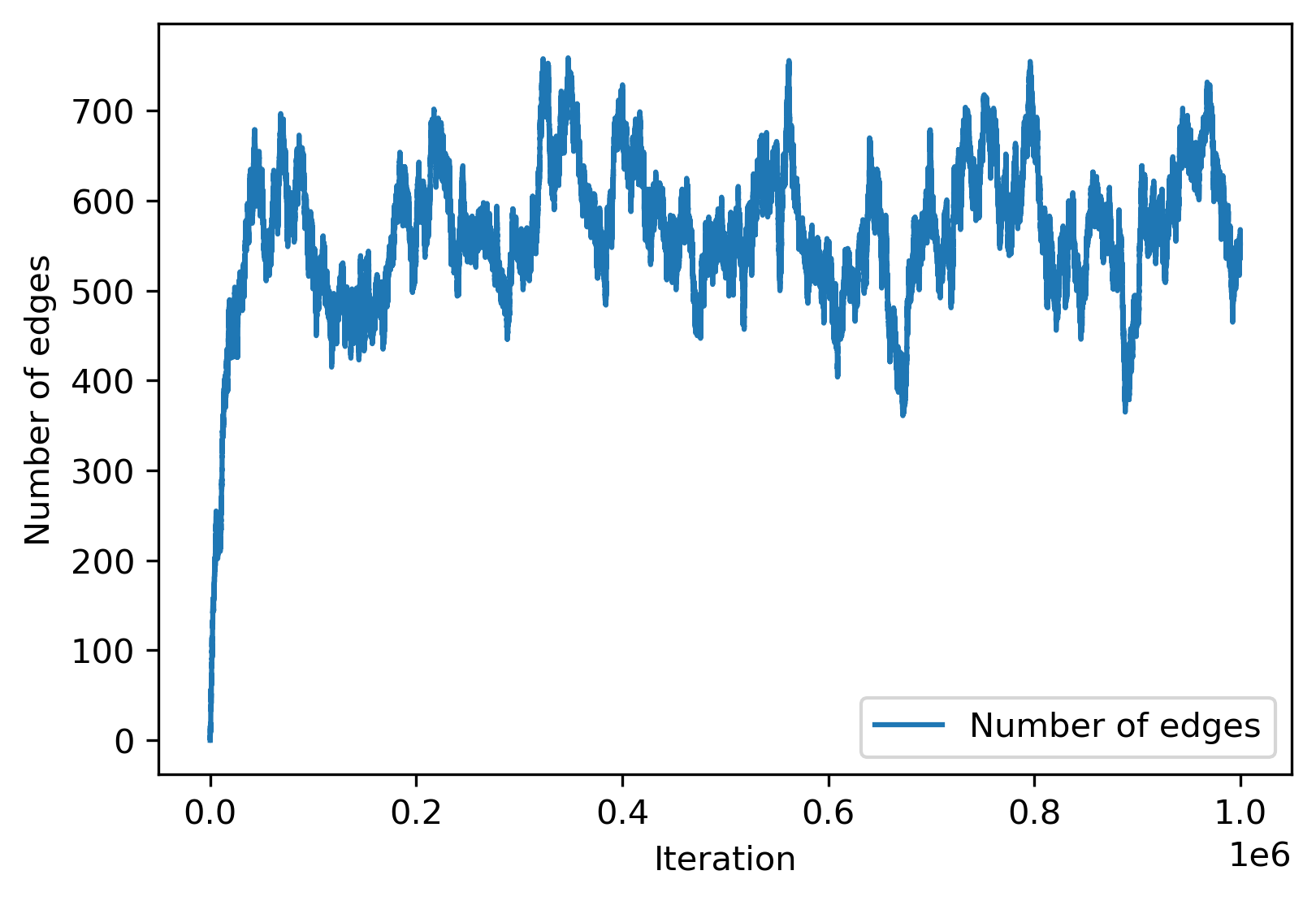}
        \caption{Double uniform}
        \label{fig:STMH_traceplot_double_uniform}
    \end{subfigure}
    \begin{subfigure}{0.45\linewidth}
        \centering
        \includegraphics[width=\linewidth]{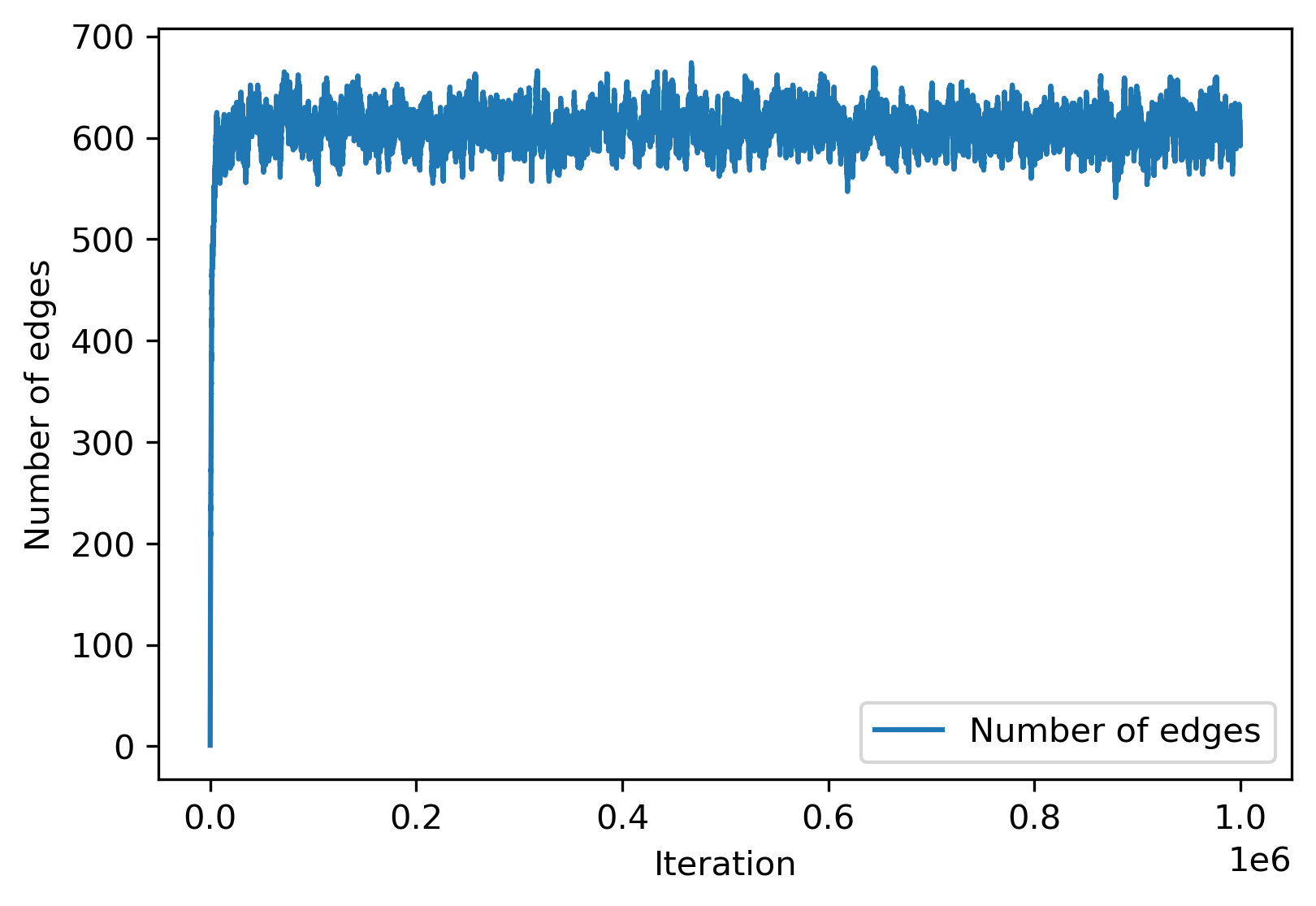}
        \caption{Uniform}
        \label{fig:STMH_traceplot_uniform}
    \end{subfigure}
     \vspace{1em}
    \begin{subfigure}{0.45\linewidth}
        \centering
    \includegraphics[width=\linewidth]{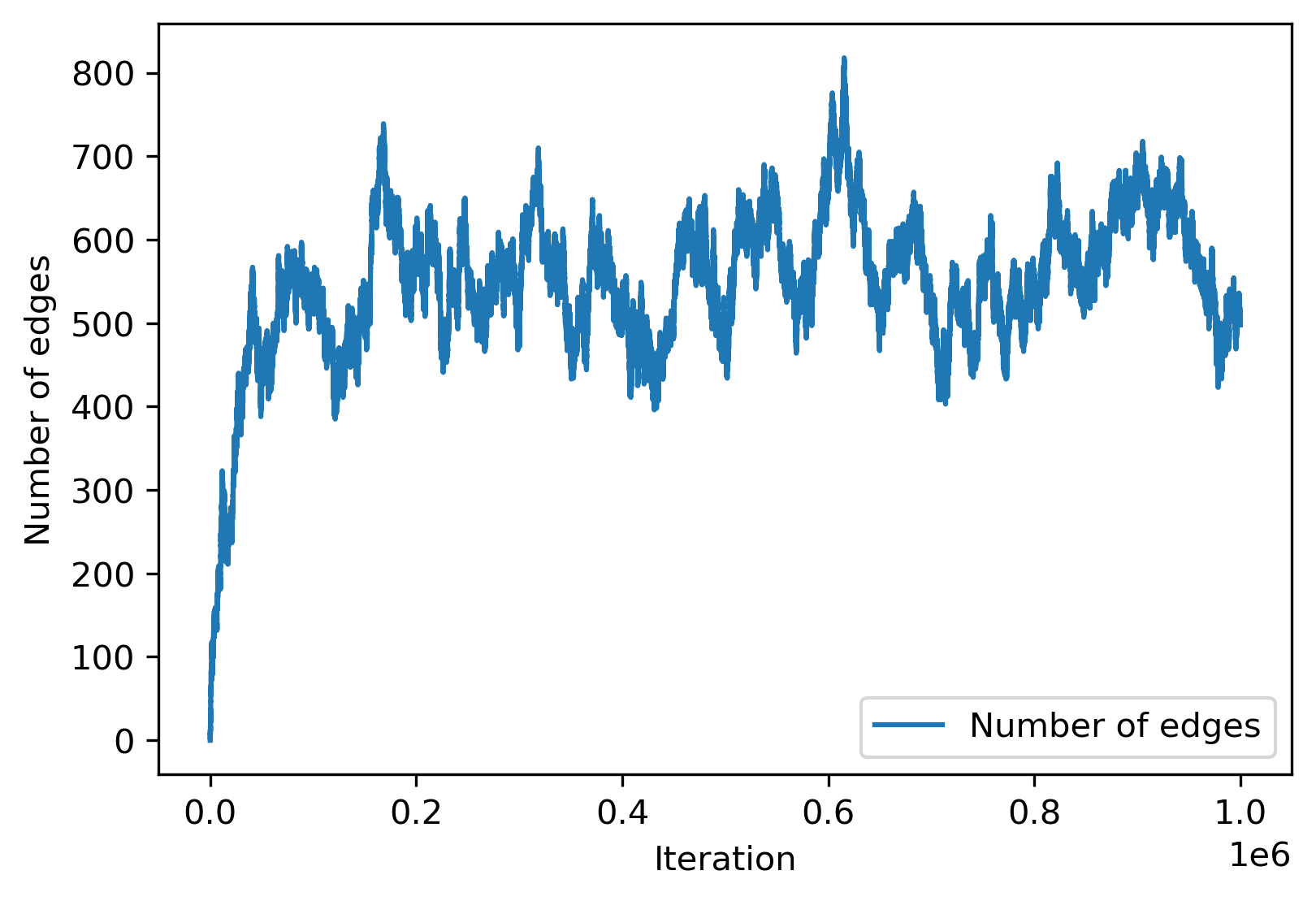}
    \caption{Truncated geometric, $\theta = 0.9901$}
    \label{fig:STMH_traceplot_truncated_geo_9901}
    \end{subfigure}
    \begin{subfigure}{0.45\linewidth}
        \centering
    \includegraphics[width=\linewidth]{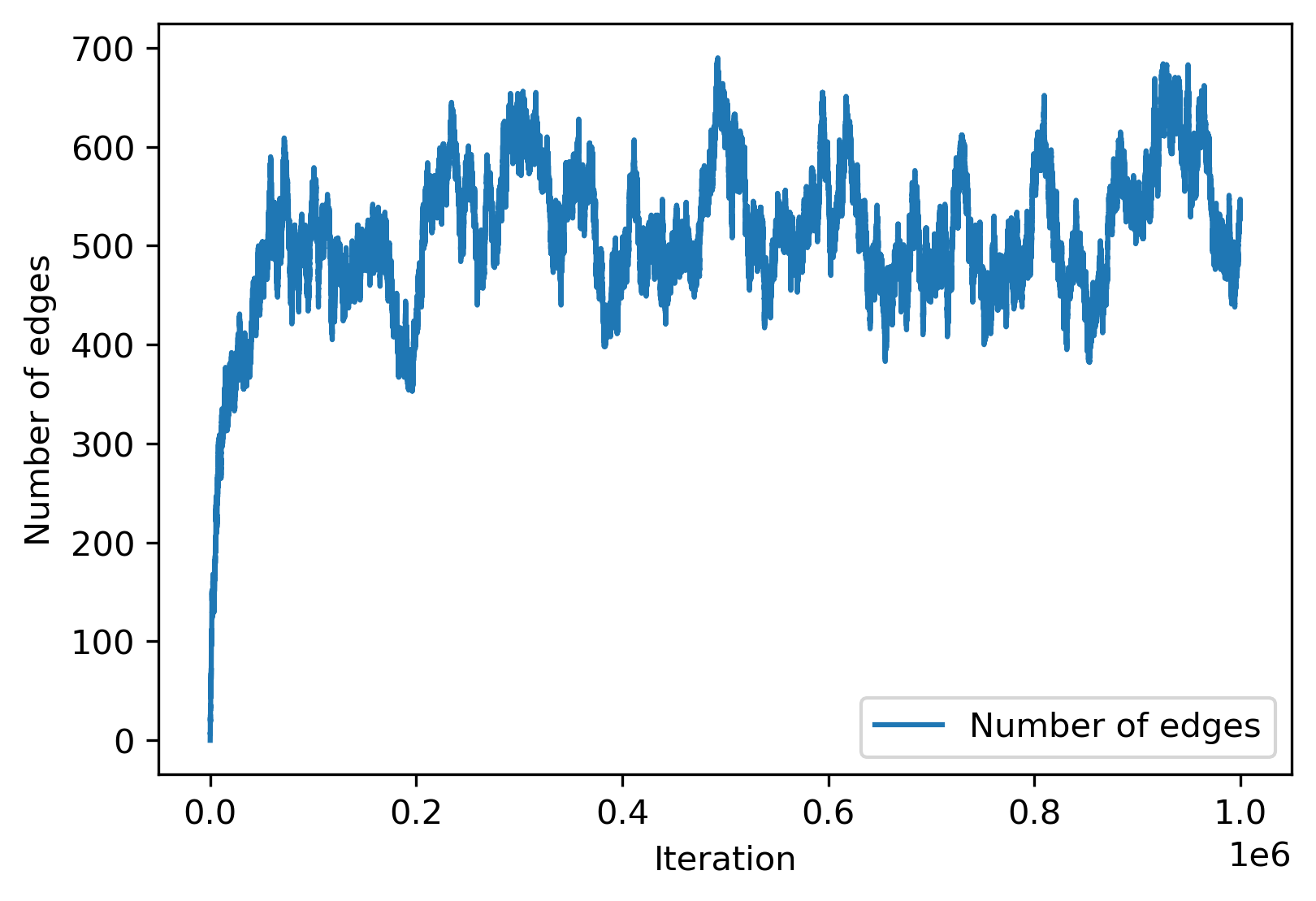}
    \caption{Truncated geometric, $\theta = 0.9804$}
    \label{fig:STMH_traceplot_truncated_geo_9804}
    \end{subfigure}
    \caption{Trace plots for posterior number of edges for the STMH algorithm when using each of the four different priors for the graph.}
    \label{fig:STMH_traceplots}
\end{figure}

\begin{figure}
    \centering
    \begin{subfigure}{0.45\linewidth}
        \centering
        \includegraphics[width=\linewidth]{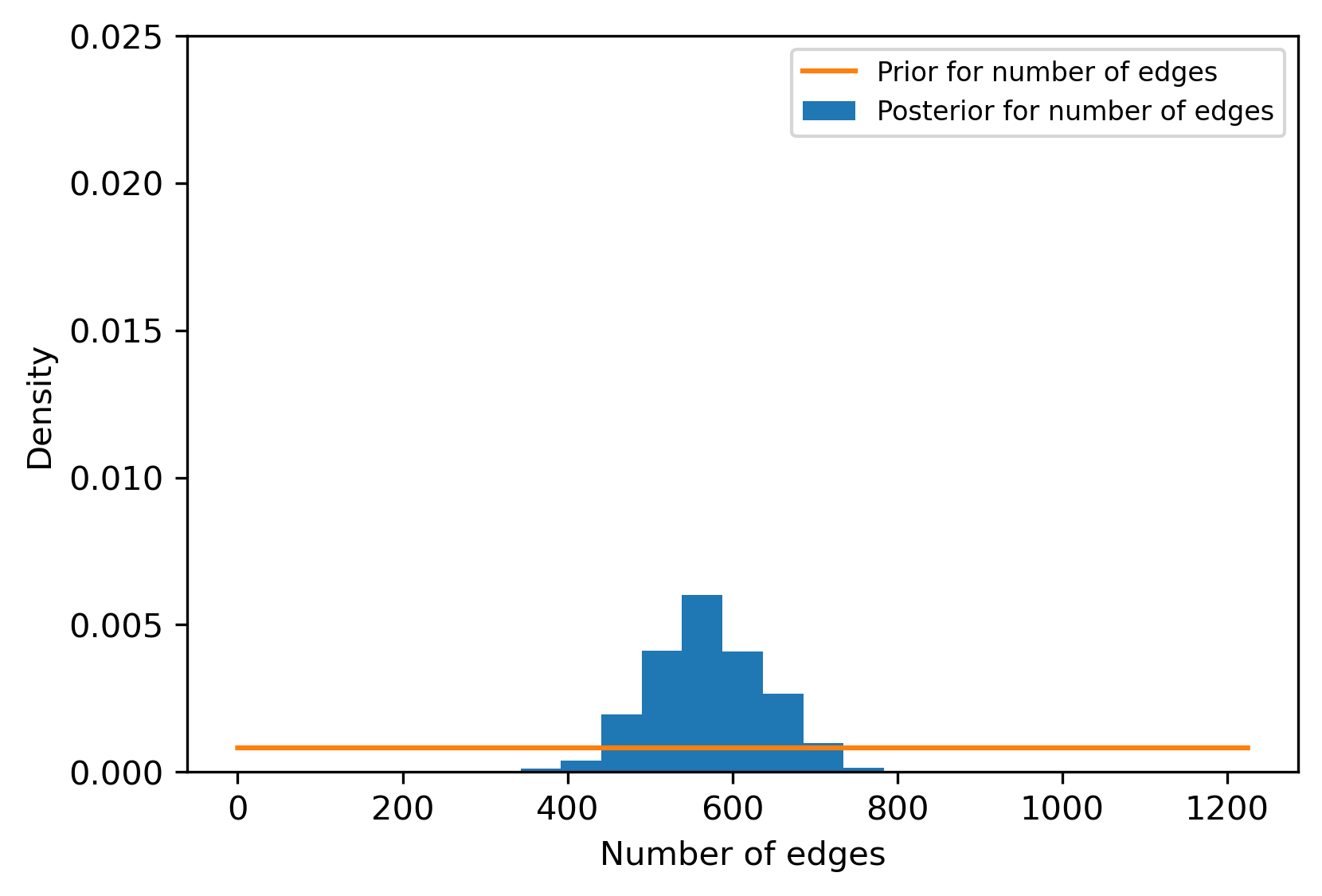}
        \caption{Double uniform}
        \label{fig:STMH_histogram_double_uniform}
    \end{subfigure}
    \begin{subfigure}{0.45\linewidth}
        \centering
         \includegraphics[width=\linewidth]{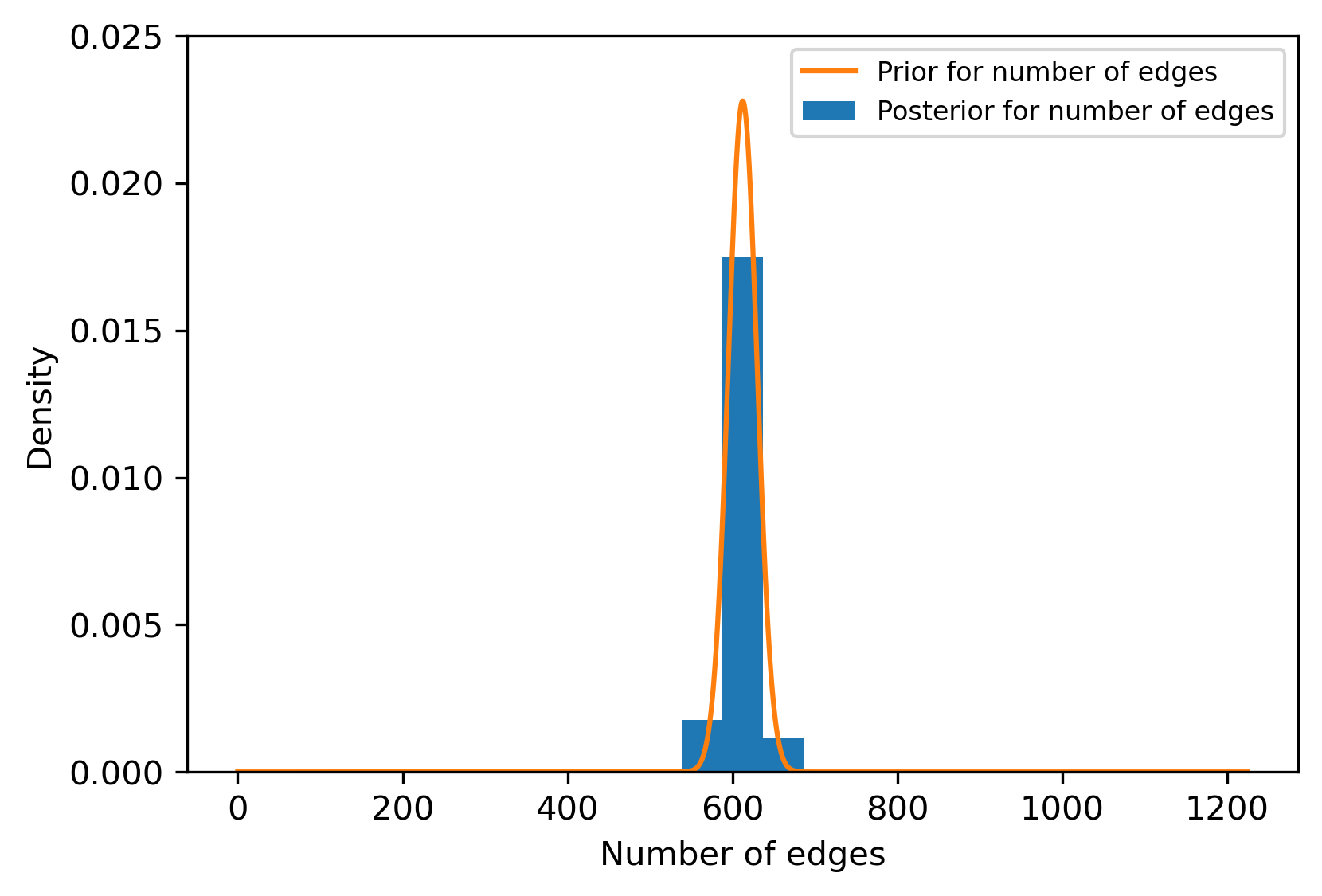}
        \caption{Uniform}
        \label{fig:STMH_histogram_uniform}
    \end{subfigure}
     \vspace{1em}
    \begin{subfigure}{0.45\linewidth}
        \centering
        \includegraphics[width=\linewidth]{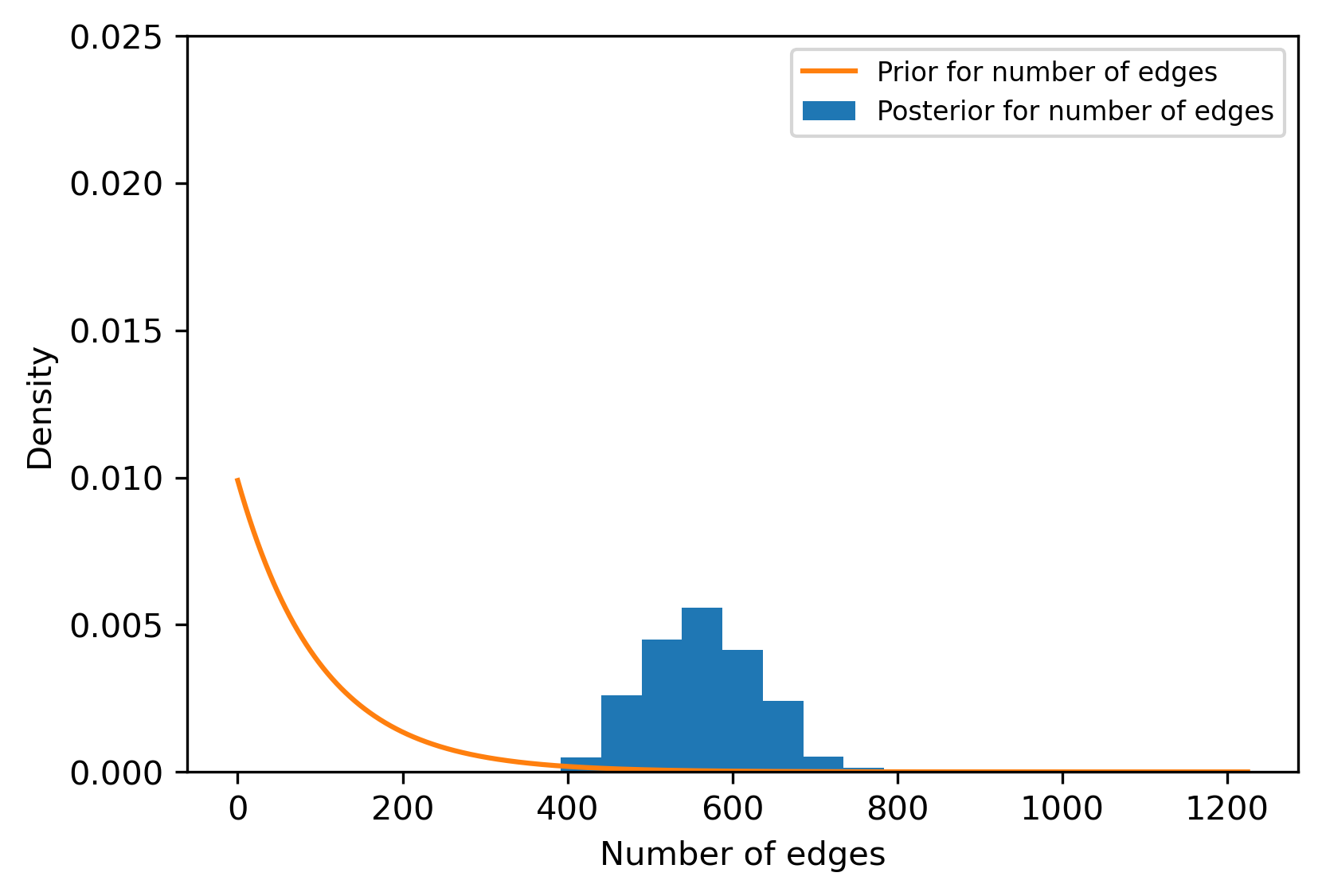}
        \caption{Truncated geometric - $\theta = 0.9901$}
        \label{fig:STMH_histogram_truncated geometric_9901}
        \end{subfigure}
        \begin{subfigure}{0.45\linewidth}
        \centering
        \includegraphics[width=\linewidth]{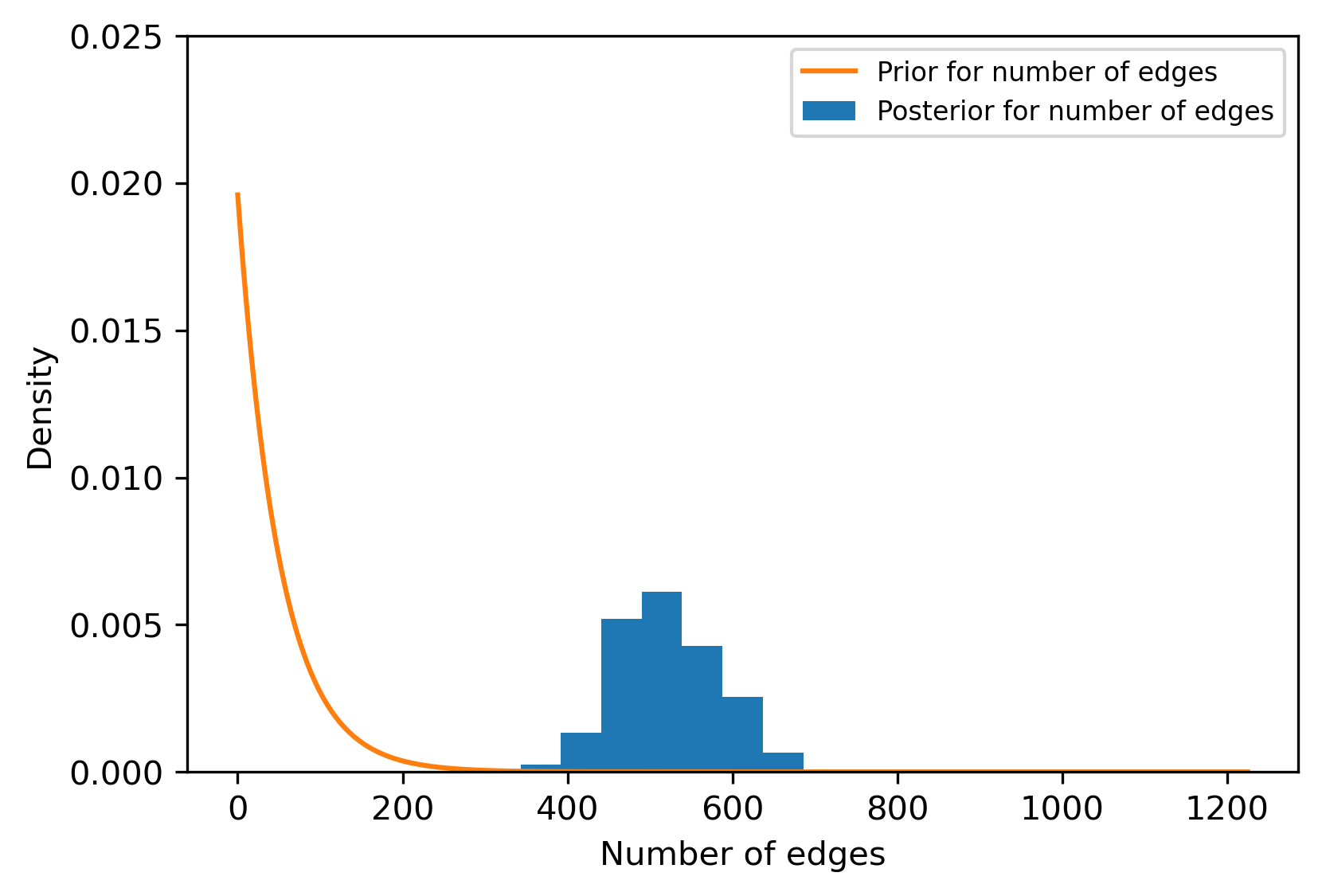}
        \caption{Truncated geometric - $\theta = 0.9804$}
        \label{fig:STMH_histogram_truncated geometric_9804}
        \end{subfigure}
    \caption{Histograms for posterior number of edges for the STMH algorithm when using each of the four different priors on the graph.}
    \label{fig:STMH_histograms}
\end{figure}

\begin{figure}
    \centering
    \begin{subfigure}{0.45\linewidth}
        \centering
        \includegraphics[width=\linewidth]{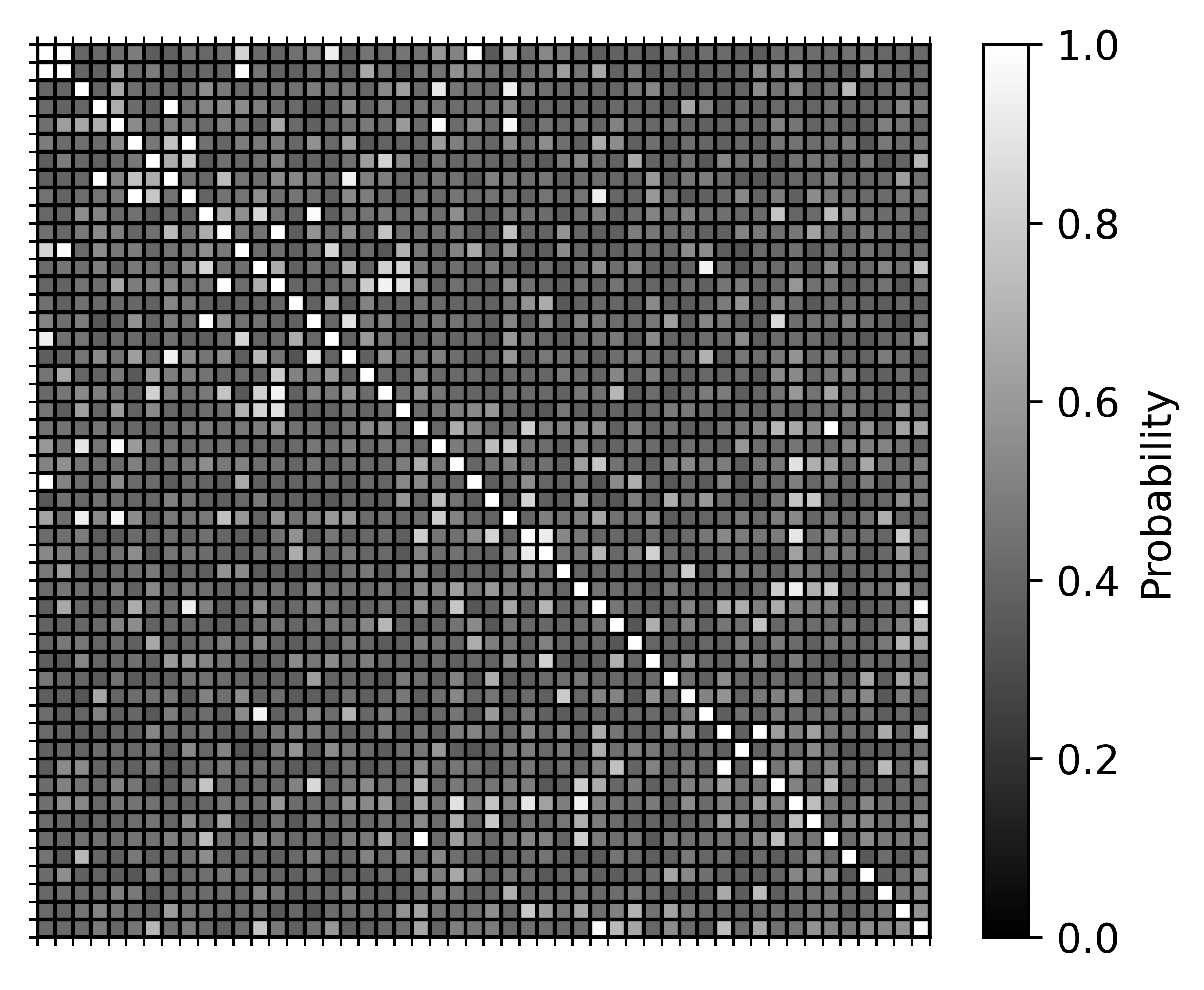}
        \caption{Double uniform}
    \label{fig:STMH_edge_prob_mat_double_uniform}
    \end{subfigure}
    \begin{subfigure}{0.45\linewidth}
        \centering
    \includegraphics[width=\linewidth]{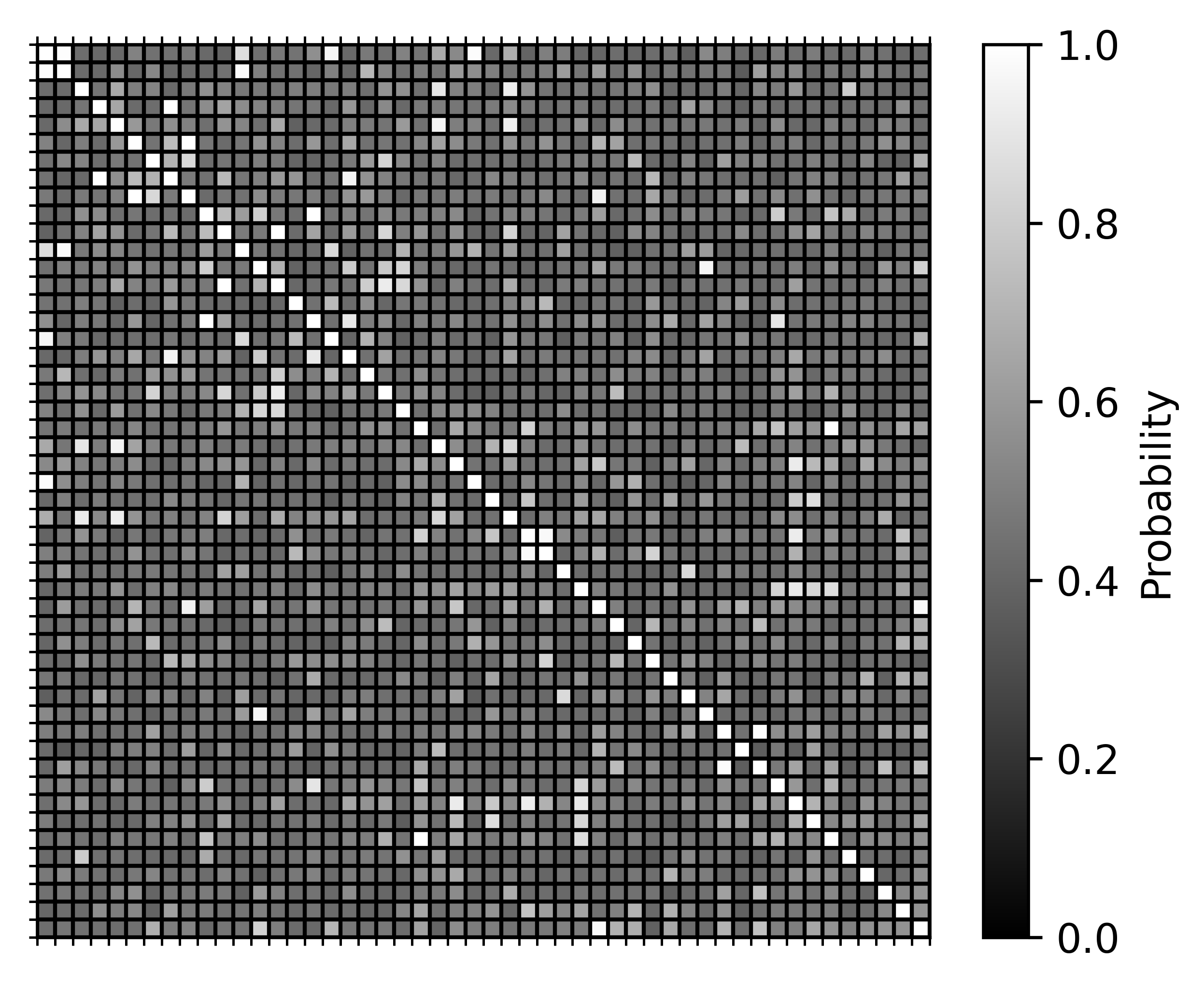}
    \caption{Uniform}
    \label{fig:STMH_edge_prob_mat_uniform}
    \end{subfigure}
     \vspace{1em}
    \begin{subfigure}{0.45\linewidth}
        \centering
    \includegraphics[width=\linewidth]{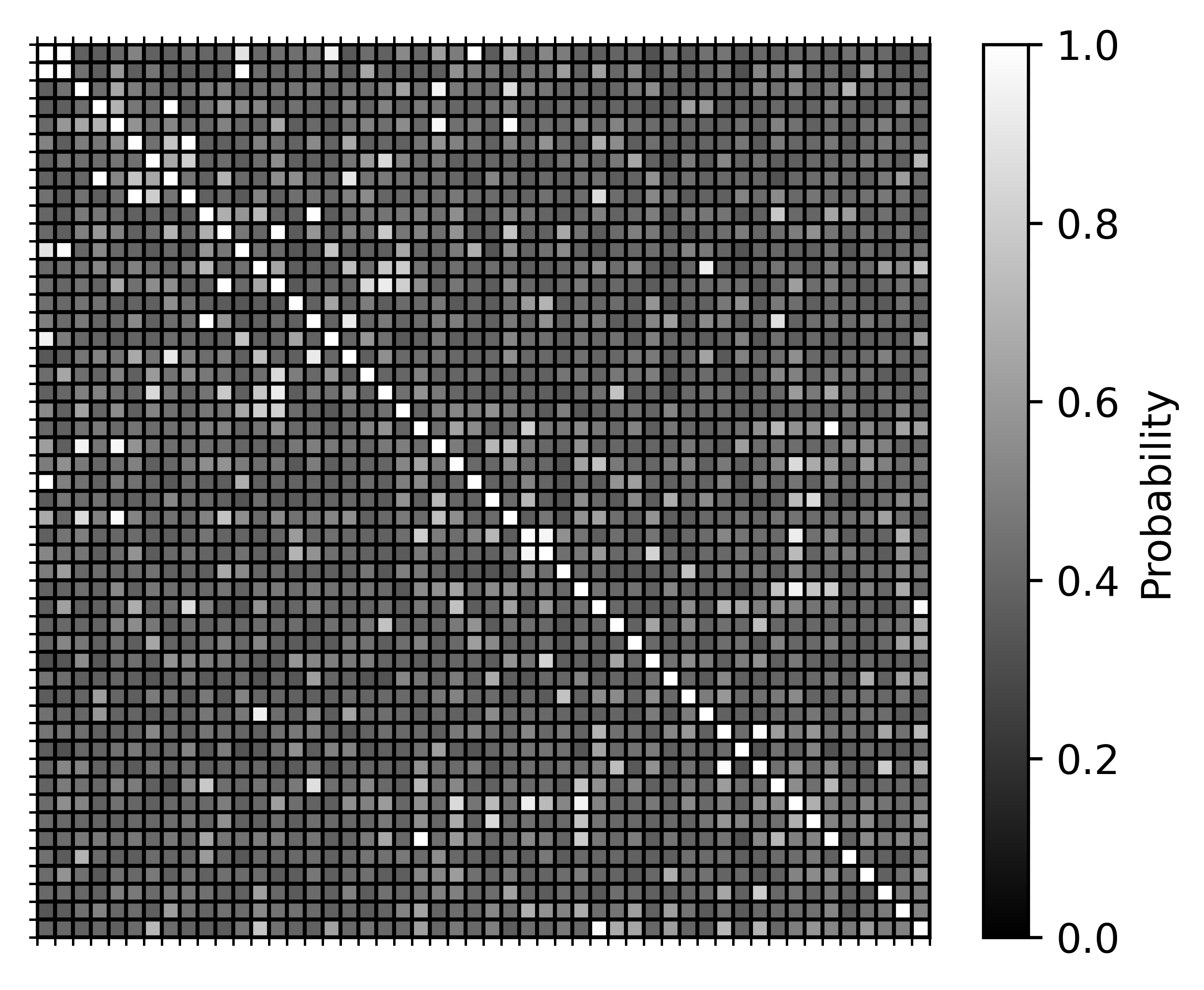}
    \caption{Truncated geometric - $\theta = 0.9901$}
    \label{fig:STMH_edge_prob_mat_truncated_geometric_9901}
        \end{subfigure}
        \begin{subfigure}{0.45\linewidth}
        \centering
    \includegraphics[width=\linewidth]{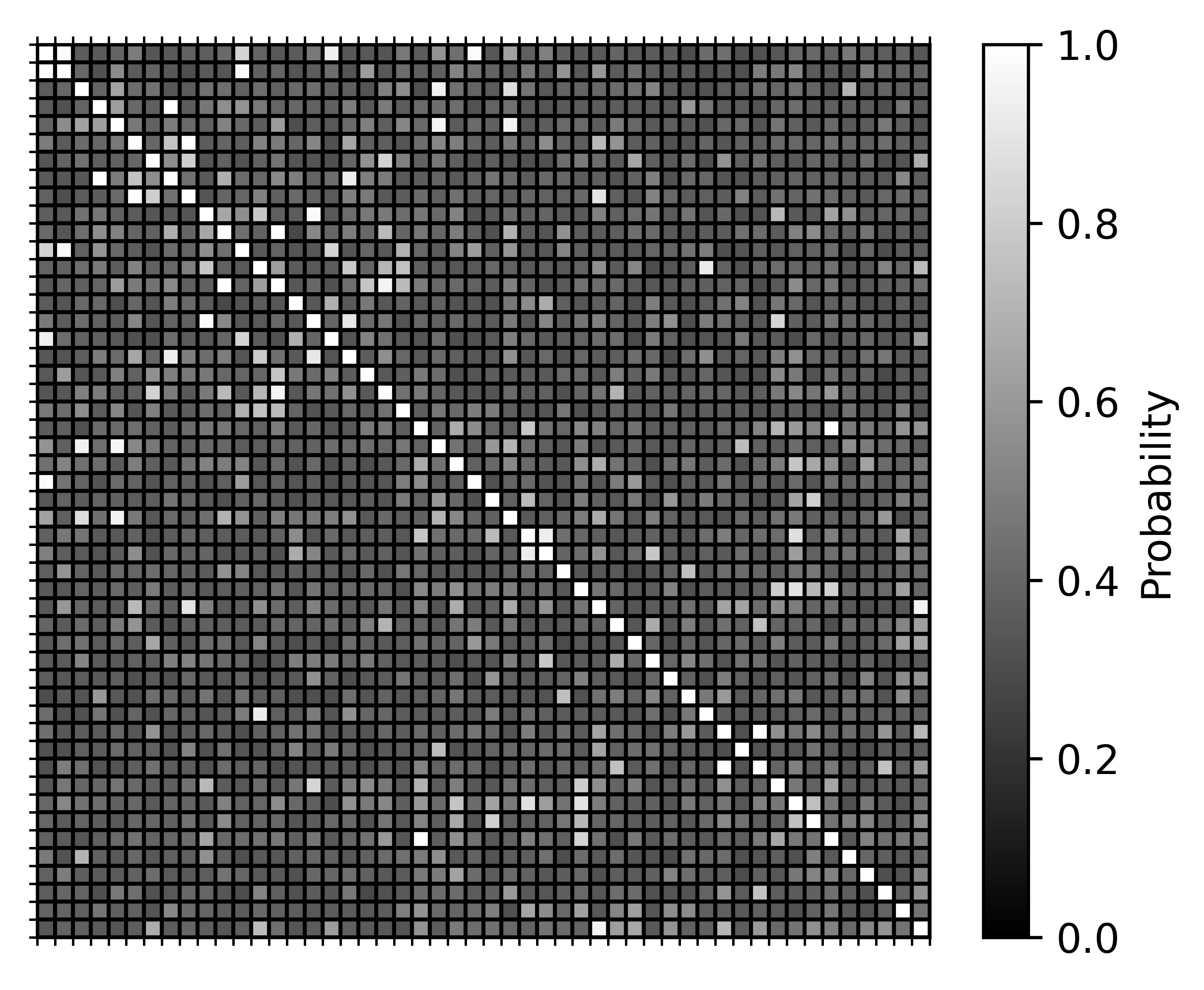}
    \caption{Truncated geometric - $\theta = 0.9804$}
    \label{fig:STMH_edge_prob_mat_truncated_geometric_9804}
        \end{subfigure}
    \caption{Estimated posterior edge probabilities for the SMTH algorithm when using each of the four different priors on the graph.}
    \label{fig:STMH_edge_prob_mats}
\end{figure}

\begin{figure}
    \centering
    \includegraphics[width=0.7\linewidth]{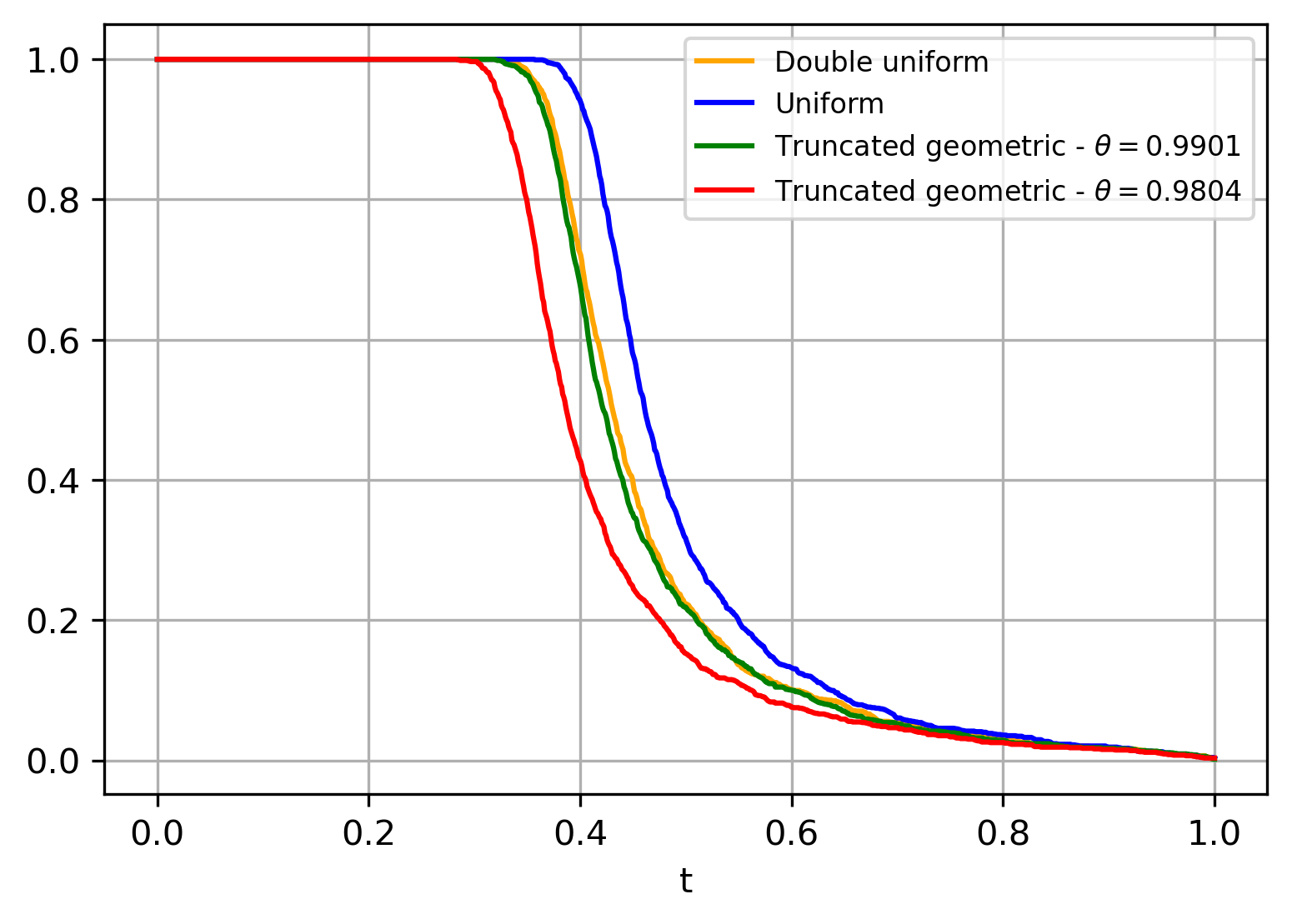}
    \caption{Fraction of edges with estimated posterior probability larger than or equal to $t$ as a function of $t$ for the SMTH algorithm with four different priors on the graph.}
    \label{fig:STMH_reverse_CDF}
\end{figure}

First, we assess the convergence and mixing of our algorithm when using each of the four different prior distributions for the graph. Figure \ref{fig:STMH_traceplots} provides trace plots for the number of edges and we can see that the number of edges seems to stabilize after about $50\,000$ iterations, which is an indication of convergence. Notably, the mixing for the uniform prior is substantially better than for the other three. Most likely, this is attributed to lower posterior variance for the number of edges, something that in turn is a result of lower prior variance for the uniform distribution than the other three. Inevitably, higher posterior variance for the number of edges naturally leads to poorer mixing, since our algorithm only can add or remove one edge at the time, something that leads to a slow exploration of the state space.
We also present some plots that highlight the differences in the posterior distributions as such. In Figure \ref{fig:STMH_histograms}, we can see histograms for the posterior number of edges when using each of the four prior distributions. Again, it is evident that the posterior associated with the uniform prior on the graph obtains much lower variance with regards to the number of edges than the other three. In addition, the mean appears somewhat higher for the posterior with the uniform prior. This is a natural consequence of the more informative nature of this choice of prior. With a uniform prior, the expected number of edges is $p(p-1)/4 = 612.5$ and since the prior variance for the number of edges is low in the uniform case, the posterior is shifted in the direction of the prior mean. In Figure \ref{fig:STMH_edge_prob_mats}, we present the estimated posterior edge probabilities when using each of the four different prior distributions, where the variables are ordered according to observed variance, from highest to lowest, before normalization. We can observe a very strong correlation between the results with the four different priors. Edges that have a high posterior edge probability for one of the priors have a high posterior edge probability for the other three as well. Still, the posterior edge probabilities are somewhat higher for the uniform case, something that aligns well with the previous discussion about more edges in the posterior for the uniform prior. Finally in Figure \ref{fig:STMH_reverse_CDF}, we provide a plot displaying the fraction of posterior edge probabilities that exceeds $t$ for an arbitrary value $t$ between zero and one.
Again, we can see that the uniform distribution exhibits higher estimated posterior edge probabilities, while the posteriors for the double uniform and the truncated geometric prior with $\theta = 0.9901$ appear to be similar with regards to this particular metric. Most likely, setting $\theta$ to $0.9901$ does not provide enough regularization to have a major effect. We can however see that the corresponding curve for the case with $\theta = 0.9804$ is shifted towards the left in relation to the others. In this case, $\theta$ is small enough to have a slight regularizing effect.

\subsection{Results for the WWA and \textit{BDgraph} algorithms}\label{subsec:res_wwa+BDgraph}
\begin{figure}
    \centering
    \begin{subfigure}{0.45\linewidth}
    \includegraphics[width=\linewidth]{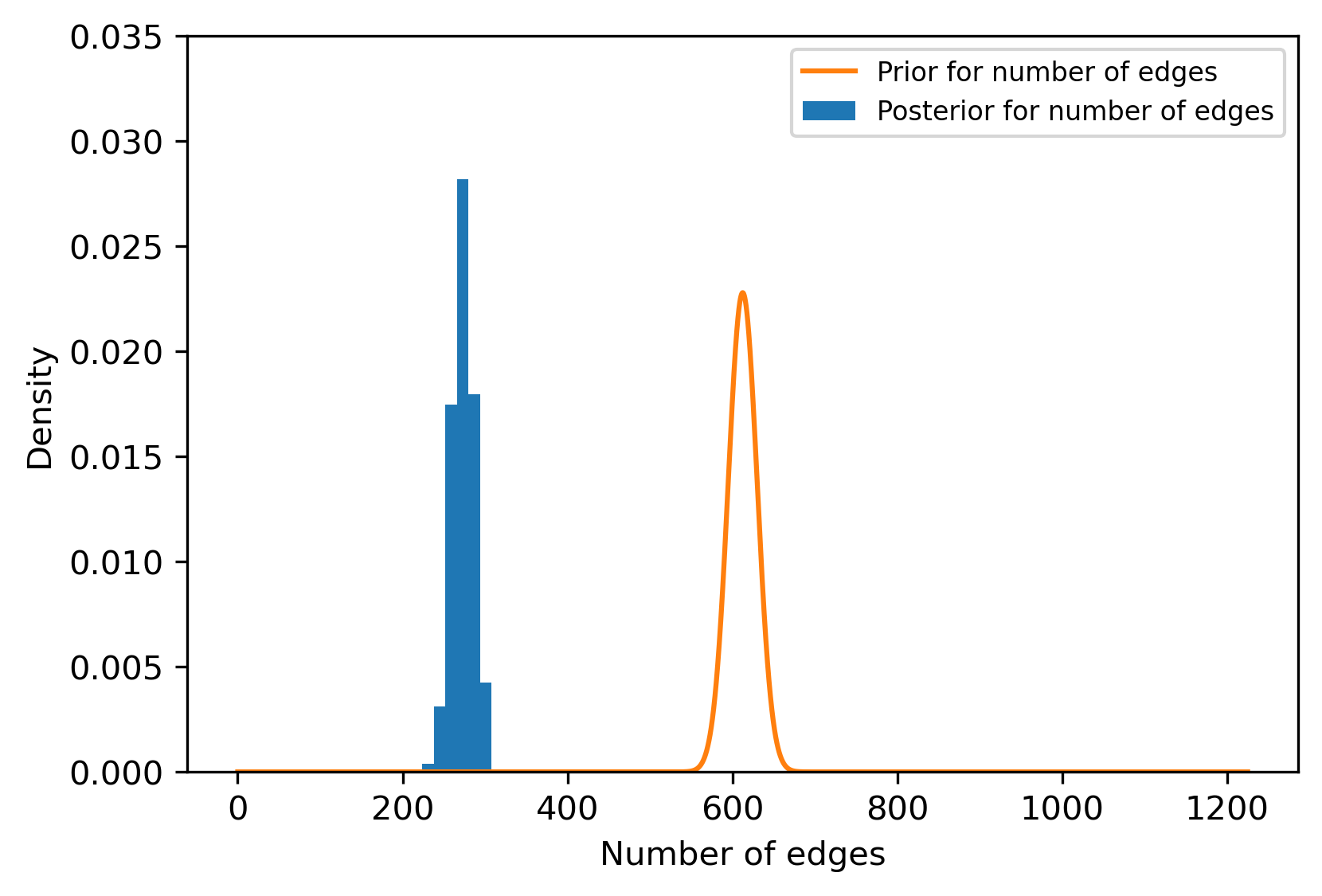}
    \caption{\textit{BDgraph} algorithm.}
    \label{fig:BDgraph_histogram}
    \end{subfigure}
    \begin{subfigure}{0.45\linewidth}
    \includegraphics[width=\linewidth]{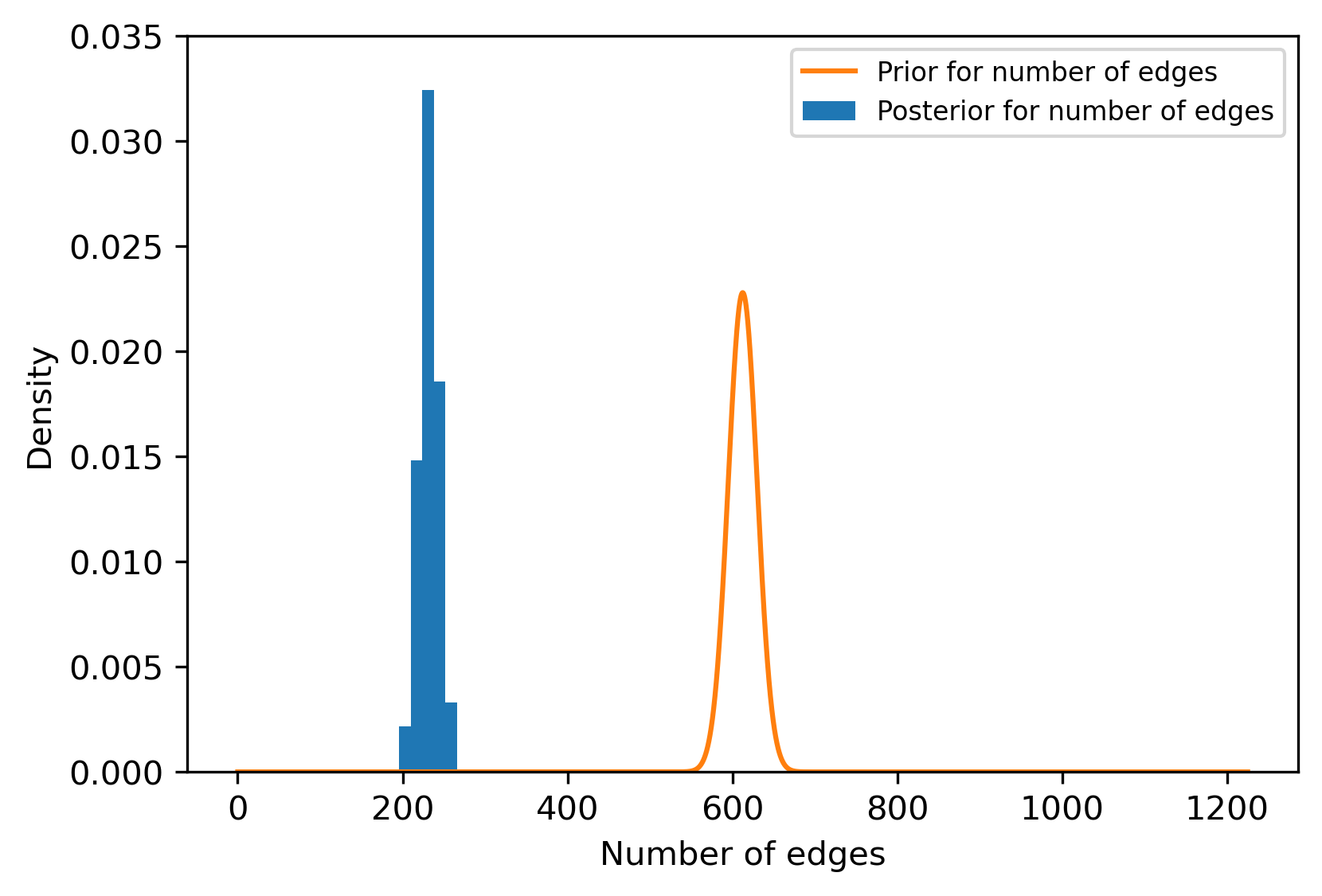}
    \caption{WWA algorithm.}
    \label{fig:WWA_histogram}
    \end{subfigure}
    \caption{Histograms for number of edges for the \textit{BDgraph} and WWA algorithms.}
    \label{fig:BDgraph+WWA_histograms}
\end{figure}

We also apply the same gene expression data set on the WWA and \textit{BDgraph} algorithms with the same pre-processing of the data as for STMH. Although \textit{BDgraph} is not an algorithm but a package, we refer to the algorithm implemented therein as simply \textit{BDgraph} in the following.
For both algorithms, we stick to the standard choice of prior for the precision matrix given by \eqref{eq:G_wishart_prior_real_data}.
Concerning the prior for the graph, the codes of both \textit{BDgraph} and WWA only offer the possibility of independent Bernoulli priors and hence, these algorithms are run with this choice with an edge probability of $0.5$. Note that this corresponds to a uniform prior on all possible graphs, which is one of the priors that was run with STMH.
Both \textit{BDgraph} and WWA are run for $50\,000$ iterations with $10\,000$ iterations as burn-in. In both algorithms, we start with an empty graph, while neither of the algorithms require initialization of $Q$ due to an initial sampling step. Histograms for the posterior number of edges for both algorithms are displayed in Figure \ref{fig:BDgraph+WWA_histograms}.

The results suggest that the posterior distribution that we obtain with the ST prior differs significantly from the ones obtained with the standard choice of G-Wishart prior (Figures \ref{fig:STMH_histogram_uniform},\ref{fig:BDgraph_histogram},\ref{fig:WWA_histogram}).
We can note that the results for the \textit{BDgraph} and WWA algorithms seem peculiar in relation to the uniform prior for the graph. For both algorithms, the posterior for the number of edges has its support far out in the tail of the prior, something that is not the case for the ST prior (Figure \ref{fig:STMH_histogram_uniform}). \citet{vanDenBoom2022} does not display the results for the case $p = 50$, but the corresponding results for $p = 100$ exhibit the same behavior, where the support of the posterior for the number of edges is very far out in the tail of the prior. 
We can see two possible explanations for this. Either, it is an effect of the precise nature of the G-Wishart prior for $Q|G$, that potentially could have the effect of significantly shifting the posterior for the number of edges for the graph in either direction. Alternatively, the effect could be a result of the incorrect sampler from \citet{LenkoskiDirect2013}, that is deployed in both the \textit{BDgraph} and WWA algorithms. 
\section{Concluding remarks}\label{sec:concluding_remarks}
In this article, we proposed a novel family of prior distributions for the precision matrix in Gaussian graphical models, called ST priors, that allow for posterior inference with MCMC without approximations of the acceptance probability. The G-Wishart distribution which has for long been the standard choice of prior, does so far not offer this possibility for larger graph sizes. 
Moreover, the family of ST prior distributions offer a lot of flexibility, since it allows us to specify the prior distribution for the free elements of the covariance matrix through the marginal of $\tilde{\pi}$.
We also proposed an MCMC algorithm for full posterior inference in a GGM for our proposed family of priors and demonstrated it on a real dataset with human gene expression data which gave satisfactory results in terms of convergence and mixing.
We also carried out inference on the same data with some standard algorithms for inference with the G-Wishart prior and compared the results.
It appears that for this dataset, that contains rather few observations in relation to the number of variables, the posterior for the number of edges is very sensitive to the choice of priors for $G$ and $Q|G$. 
One can note that for a Wishart distribution, unlike an Inverse Wishart distribution, there is a limit for how large marginal variance we can obtain for the elements with fixed expectation. For this reason, our choice of prior can be regarded as more informative than the one in \eqref{eq:G_wishart_prior_real_data} and this is something that could be interesting to examine further.

One possible extension of the ST priors proposed here, would be to retain a prior for $\Sigma$, $\tilde{\pi}$, with support in $\mathbb{P}$ but to redefine the prior for $Q|G$ through
\begin{equation*}
    Q = \widehat{\text{PD}}_G(\Sigma),
\end{equation*}
where $\widehat{\text{PD}}_G(\cdot)$ corresponds to running a PD-completion algorithm for a fixed number of iterations or with some error tolerance larger than zero. The aim of this would be computational speedup, with the drawback of sacrificing some precise knowledge of what the prior actually is. Note that an approach of this kind would require the use of the IPS algorithm, since aborting the Hastie algorithm prematurely, would not guarantee the correct sparsity pattern, since it operates on $Q^{-1}$ rather than $Q$.

Another aspect that could be further investigated is the design of MCMC algorithms for the ST prior family with better proposal distributions. One such possibility would be to propose joint updates of graph and $\Sigma$, for instance by letting the value of $\Sigma_{ij}$, when proposing to add the edge $(i,j)$, be informed by the data. \citet{vanDenBoom2022} made use of informed proposals for the graph by exploiting approximations of the normalizing constant for the G-Wishart distribution. Such approximations are not readily available in the context of ST priors, but one possibility would be to examine an approximation to this construction in our setting. Another possible extension would be to relax the assumption of independence between $G$ and $\Sigma$ in \eqref{eq:joint_G_Sigma_prior} such that $\tilde{\pi}$ in the ST prior for $Q|G$ depends on the graph. The aim would be to provide more flexibility. However, when $\Sigma$ and $G$ are no longer independent, we need the normalizing constant of $\tilde{\pi}$ in order to carry out MCMC inference without approximations of the acceptance probability.
\section*{Disclosure Statement}
The authors report there are no competing interests to declare.
\section*{Funding}
This work was supported by the SFI Centre for Geophysical Forecasting, (Norwegian Research Council grant no. 309960).
\section*{Data availability statement}
The data used in this article is publicly available through the \textit{BDgraph} package (DOI: 10.32614/CRAN.package.BDgraph).
\bibliography{main.bib}
\end{document}